\documentclass[a4paper,12pt]{article}
\usepackage{amssymb}
\usepackage{amsfonts}
\usepackage{amsmath}
\usepackage{amstext}
\usepackage{epsfig}
\usepackage{dcolumn}
\usepackage{rotating}
\usepackage{color}
\usepackage{comment}
\usepackage{hyperref}

\def\Xint#1{\mathchoice
{\XXint\displaystyle\textstyle{#1}}%
{\XXint\textstyle\scriptstyle{#1}}%
{\XXint\scriptstyle\scriptscriptstyle{#1}}%
{\XXint\scriptscriptstyle\scriptscriptstyle{#1}}%
\!\int}
\def\XXint#1#2#3{{\setbox0=\hbox{$#1{#2#3}{\int}$ }
\vcenter{\hbox{$#2#3$ }}\kern-.56\wd0}}

\def\dashint{\Xint-}

\newcommand*\xbar[1]{%
  \hbox{%
    \vbox{%
      \hrule height 0.5pt % The actual bar
      \kern0.5ex%         % Distance between bar and symbol
      \hbox{%
        \kern-0.1em%      % Shortening on the left side
        \ensuremath{#1}%
        \kern-0.1em%      % Shortening on the right side
      }%
    }%
  }%
}

\definecolor{rosso}{cmyk}{0,1,1,0.4}
\definecolor{rossos}{cmyk}{0,1,1,0.55}
\definecolor{rossoc}{cmyk}{0,1,1,0.2}
\definecolor{blu}{cmyk}{1,1,0,0.3}
\definecolor{blus}{cmyk}{1,1,0,0.6}
\definecolor{bluc}{cmyk}{1,1,0,0.1}
\definecolor{verde}{cmyk}{0.92,0,0.59,0.25}
\definecolor{verdec}{cmyk}{0.92,0,0.59,0.15}
\definecolor{verdes}{cmyk}{0.92,0,0.59,0.7}

\newcommand{\ba}{\begin{eqnarray}}
\newcommand{\ea}{\end{eqnarray}}
\newcommand{\be}{\begin{equation}}
\newcommand{\ee}{\end{equation}}
\newcommand{\bi}{\begin{itemize}}
\newcommand{\ei}{\end{itemize}}
\newcommand{\al}{\alpha}
\newcommand{\bt}{\beta}
\newcommand{\ga}{\gamma}

\newcommand{\da}{\delta}
\newcommand{\la}{\lambda}

\newcommand{\sa}{\sigma}
\newcommand{\en}{\epsilon}

\newcommand{\Ga}{\Gamma}

\newcommand{\La}{\Lambda}
\newcommand{\cF}{{\cal F}}

\newcommand{\cK}{{\cal K}}
\newcommand{\cL}{{\cal L}}

\newcommand{\cO}{{\cal O}}

\newcommand{\w}{\widetilde}

\newcommand{\st}{\stackrel}

\newcommand{\p}{\partial}
\newcommand{\hp}{h^{\perp}}
\newcommand{\n}{\nabla}

\newcommand{\ra}{\rightarrow}
\newcommand{\Ra}{\Rightarrow}

\newcommand{\LF}{\left(}
\newcommand{\RF}{\right)}
\newcommand{\LT}{\left[}
\newcommand{\RT}{\right]}
\newcommand{\Ld}{\left.}
\newcommand{\Rd}{\right.}
\newcommand{\kb}{\bar{k}}
\newcommand{\pb}{\bar{p}}
\newcommand{\4}{\frac{1}{4}}

\newcommand{\mx}{\mbox}
\newcommand{\mt}{\mathtt}

\newcommand{\where}{\mx{ where }}

\newcommand{\ie}{{\it i.e.\ }}

\newcommand{\non}{\nonumber\\}
\begin{document}

\tolerance=100000
\thispagestyle{empty}
\vspace{1cm}

\begin{center}
{\LARGE \bf
Towards understanding the ultraviolet behavior of quantum loops in  infinite-derivative theories of gravity
 \\ [0.15cm]
% \&
% \\ [0.18cm]
% when it is long
}
\vskip 2cm
{\large Spyridon Talaganis$^a$, Tirthabir Biswas$^b$ and Anupam Mazumdar$^{a,~c}$}
\vskip 7mm
{\it $^a$ Consortium for Fundamental Physics, Physics Department, Lancaster University, \\ Lancaster, LA1 4YB, UK}
\vskip 3mm
{\it $^b$ Department of Physics, Loyola University, \\ 6363 St. Charles Avenue, Box 92, \\ New Orleans, LA 70118, USA}
\vskip 3mm
{\it $^c$ D\'{e}partement de Physique Th\'eorique, Universit\'e de Gen\'eve, 24, Quai E Ansermet, 1211 Gen\'eve 4, Switzerland}
\end{center}
\date{\today}

\begin{abstract}
In this paper we will consider quantum aspects of a non-local, infinite-derivative scalar field theory - a {\it toy model} depiction of a covariant infinite-derivative, non-local extension of Einstein's general relativity which has previously been shown to be free from ghosts around the  Minkowski background.  The graviton propagator in this theory gets an exponential suppression making it {\it asymptotically free}, thus providing strong prospects of resolving various classical and quantum divergences. In particular, we will find that at $1$-loop, the $2$-point function is still divergent, but once this amplitude is renormalized by adding appropriate counter terms, the ultraviolet (UV) behavior of all other $1$-loop diagrams as well as the $2$-loop, $2$-point function  remains well under control.  We will go on to discuss how one may be able to generalize our computations and arguments to arbitrary loops.
\end{abstract}

\newpage

\setcounter{page}{1}

\tableofcontents
%%%%%%%%%%%%%%%%%%%%%%%%%%%%%%%%%%%%%%%%%%%%%%%%%%%%%
\section{Introduction}
\numberwithin{equation}{section}

Formulating a quantum theory of gravity~\cite{Veltman:1975vx,dewittQG,DeWitt:2007mi} remains one of the most outstanding challenges of high energy physics. While string theory (ST)~\cite{Polchinski:1998rr} remains the most popular candidate, other notable efforts include Loop Quantum Gravity (LQG)~\cite{Ashtekar,Nicolai:2005mc}, Causal Set approach~\cite{Henson:2006kf}, and ideas based on asymptotic safety~\cite{Weinberg:1980gg}. An interesting recurrent feature that appears in several of these approaches is non-locality. For instance, the entire formulation of LQG is based on non-local objects, such as Wilson loops and fluxes coming from the gravitational field.  Strings and branes of ST are, by their very definition, non-local objects. Even classically they do not interact with each other at a specific spatial point, but rather over a region in space. Not surprisingly, non-local structures are a common theme in stringy field theory (SFT) models. For instance, these appear in noncommutative geometry~\cite{noncom} \& SFT~\cite{Witten:1985cc},  for a review, see \cite{Siegel:1988yz}, and various {\it toy model}s of SFT such as $p$-adic strings~\cite{Freund:1987kt}, zeta strings~\cite{Dragovich:2007wb}, and strings quantized on a random lattice~\cite{Douglas:1989ve,Biswas:2004qu}. A key feature of these models is the presence of an {\it  infinite series of higher-derivative} terms incorporating the non-locality in the form of an {\it exponential kinetic} correction. Finally, it is also intriguing to note that similar infinite-derivative modifications have also been argued to arise in the asymptotic safety approach to quantum gravity~\cite{Krasnov}.

%~\cite{Tomboulis:1980bs,Tomboulis:1983sz,Tomboulis:1997gg,Biswas:2011ar,Biswas:2013kla,Modesto:2011kw,BG,Barvinsky:2014lja}.
%For a review on these topics, see~\cite{BT}.

Accordingly, in \cite{Siegel:2003vt,Tseytlin:1995uq,Biswas:2005qr} attempts were made to construct ghost-free, infinite-derivative theories of gravity which may be able to resolve space-time singularities such as the ones present  inside the black holes and at the big bang. For instance, in~\cite{Biswas:2005qr} a non-singular bouncing
cosmological background was obtained within a class of infinite-derivative gravity theories, around which the sub and super-Hubble perturbations are well behaved and do not show
instabilities~\cite{Biswas:2010zk,Biswas:2012bp,Craps:2014wga}. In fact, such an action can also modify the famous Raychaudhuri's equation and alter the Hawking-Penrose singularity theorem~\cite{Conroy:2014dja}, which can yield a non-singular bouncing cosmology without violating the null energy conditions.

It was not until recently though, that concrete criteria for any covariant gravitational theory (including infinite-derivative theories) to be free from ghosts and tachyons around the Minkowski vacuum was obtained  by Biswas, Gerwick, Koivisto and Mazumdar (BGKM)~\cite{Biswas:2011ar,Biswas:2013kla}; see~\cite{Eliezer:1989cr} for a detailed exposition of the problem of instabilities in infinite-derivative theories. In Ref.~\cite{Biswas:2011ar}, it was also shown how one can construct infinite-derivative theories of gravity where no new perturbative states are introduced and only the graviton propagator is modified by a multiplicative entire function. In particular, one can choose the entire function to correspond to the gaussian which suppresses the ultraviolet (UV) modes making the theory asymptotically free. For brevity we will refer to this case  as the BGKM model.

Given the prospects of the BGKM model at resolving the classical singularities of GR, see \cite{BT} for an overview, here we are going to explore the possibility of formulating a quantum theory of BGKM, and  the various challenges we need to overcome. For important works on slightly different approaches to quantizing gravity involving infinite-derivative interactions, see~\footnote{Regarding the differences between the ``BGKM'' model and Refs.~\cite{Tomboulis,Modesto}, Ref.~\cite{Tomboulis,Modesto} uses propagators that go as $k^{-2\ga-4}$, $\ga \geq 2$, in the UV while our propagator falls off exponentially (the exponential fall-off in the propagator is also seen as a special case in Ref~\cite{Modesto} and in Ref.~\cite{Calcagni:2014vxa}). In particular, this changes the degree of divergence, which, in our case, is a modified one counting not powers of momenta but exponents, and the divergence structure. Furthermore, we had to develop new techniques for regulating and evaluating the Feynman integrals. Also, the loop integrals are computed explicitly in our work. The ``BGKM'' model has also been shown to address cosmological and black-hole singularities, as~\cite{BigBang,Calcagni:2014vxa} also do. Hence, we decided to give these theories a different name to distinguish them from other non-local/infinite-derivative models in vogue.}~\cite{Tomboulis,Modesto,BG,Anslemi,Moffat-qg,Addazi}.
It's probably worth mentioning that in recent years there has been a growing interest in infinite-derivative gravitational theories in not only addressing the Big bang singularity problem~\cite{Biswas:2005qr,Biswas:2010zk,Biswas:2012bp,Craps:2014wga,BigBang} but also finding other cosmological applications~\cite{cosmology} and the gravitational entropy~\cite{entropy}.

Let us start by  recalling the canonical examples of infinite-derivative actions that appear in string literature. These can all be written as
\be
S=\int d^{D}x \, \LT\frac{1}{2}\phi \cK(\Box)\phi -V_{\mt{int}}(\phi)\RT\,,
\ee
where the kinetic operator $\cK(\Box)$ contains an infinite series of higher-derivative terms.
For instance, we find that $\cK(\Box)=-e^{\Box/M^2}$ for stringy {\it toy model}s based on $p$-adic numbers~\cite{Freund:1987kt}, or random lattices~\cite{Douglas:1989ve,Biswas:2004qu}, and $\cK(\Box)=-(\Box+m^2)e^{\Box/M^2}$  in SFT~\cite{Witten:1985cc}, where $m^2(<0)$ and $M^2(>0)$ are proportional to the string tension~\footnote{Here and hereafter, we are going to use  $(-+++)$ as our metric signature convention.}. Apart from its stringy origin, the above theories are interesting in their own right. Firstly, although these theories contain higher derivatives, they do not contain ghosts, at least perturbative. To see this explicitly, one can consider a fourth-order scalar theory with $\cK(\Box)=-\Box(1+\frac{\Box}{m^2})$. The corresponding propagator reads
\be
\Pi(p^2)\sim \frac{1}{p^2(p^2 -m^2)}\sim \frac{1}{p^2 -m^2}-\frac{1}{p^2}\,.
\ee
From the pole structure of the propagator it is clear that the theory contains two physical states, but unfortunately the massive state has the ``wrong'' sign for the residue indicating that it is a ghost. Once interactions are included, it makes the classical theory unstable, and the quantum theory non-unitary (see Refs.~\cite{Efimov,Tomboulis:1983sw} regarding the issue of unitarity in infinite-derivative theories). The stringy kinetic modifications combine to be an exponential, which is an {\it entire function} without any zeroes. In other words, it does not introduce any new states, ghosts or otherwise. Indeed, this property has been exploited to construct various non-local infinite-derivative theory and particle phenomenology models~\cite{Biswas:2014yia,Biswas:2009nx,Moffat:1990jj,Bluhm,Reddy}, and scalar field cosmology with infinite derivatives~\cite{scalar-cosmology}.

Secondly, as mentioned before, the infinite-derivative modification preserves a well known property of higher-derivative theories, that of making the quantum loop contributions better behaved in the UV. The stringy infinite-derivative scalar theories not only ameliorates the UV behavior, but the exponential suppression in the propagator actually makes all the quantum loops finite. Such calculations were used to provide evidence for several  stringy phenomena, such as Regge behavior~\cite{Biswas:2004qu} and thermal duality and Hagedorn transition~\cite{Biswas:2009nx}. It is then natural to wonder whether such non-local features can help in solving the quantum UV problem of gravity? In fact, Stelle, in Ref.~\cite{Stelle:1976gc} argued that the simplest higher-derivative theory of gravity, namely the fourth-order theory is already renormalizable, see also~\cite{Goroff:1985th,Goroff:1985sz}. Unfortunately,  the theory contains ghosts and is non-unitary. In contrast, the BGKM model provides  gravitational analogues of Eq.~\eqref{nlaction} where  the graviton propagator obtains an additional exponential suppression just as  the scalar models.

So, can this exponential infinite-derivative modification also solve the quantum UV problem of gravity by making all the Feynman loops finite? The answer is not  straightforward and our paper is essentially an {\it effort} to address this question. The main problem with the gravitational theories, as opposed to a scalar field theory, is that it is a gauge theory. And, one of the key features of gauge theories is that its free kinetic action is related to the interaction terms via the gauge symmetry. We will see that the exponential suppressions in propagators inevitably give rise to exponential enhancements in the vertex factors. Actually, this compensating interplay between propagators and vertices is not unique to infinite derivative theories, but any covariant theory of gravity, including Einstein's theory and Stelle's $4$th order gravity~\cite{Stelle:1976gc}, see also~\cite{Goroff:1985th,Goroff:1985sz}. In particular, the compensation between propagators and vertices is exact at the $1$-loop level making these contributions divergent as in GR. However, for higher loops, the superficial degree of divergence calculations is different from GR, because the counting is based on the pre-factors of the exponents rather than the degree of polynomial divergence; exponentials dominate any polynomial growth in the UV. In fact, a naive superficial divergence counting does suggest that diagrams with more than one loop  should be finite~\cite{Tomboulis,Modesto}.

The principal aim of this paper is to investigate the validity of such divergence counting in some details in a simplified {\it toy model} which retains the compensating feature of exponential suppression and enhancements between the propagator and interaction respectively. We will consider a  scalar field action, which maintains a combination of global scaling and  shift symmetries, similar to the residual
symmetry of gravity around the Minkowski background. Although this symmetry manifests itself only at the level of  classical equations of motion, it will allow us to incorporate the compensating feature of exponential suppression and enhancement in propagators and interactions, respectively, that is present in the full gravitational theory. We found that one nice way  to introduce this opposing effect in scalar models is to invoke the scaling symmetry. This is not central to our discussion, but rather than invoking an action in a completely ad hoc fashion, we felt that this gives us a slightly better motivation.

We will consider a cubic interaction that respects the symmetry  and study $2$- and higher-point functions at $1$- and $2$-loops. We will first look at the vanishing external momentum limit, as they are technically easier to analyse and can already tell us whether a graph will be finite or not. It should be emphasized that we had to develop new techniques for regulating and evaluating the Feynman integrals. We will next look at the finite external momentum case, which is important in determining whether renormalizability arguments can be recursively pursued or not. Although, the cubic scalar interaction
inherits a bad IR behavior - being unbounded from below, still it serves as a very good example to study the UV aspects of the theory, which is the main focus of our paper. In particular, we will employ both hard cutoff and dimensional regularization techniques to regulate the loop integrals, and we will speculate how higher loops in these theories may also remain finite in the UV.

The paper is organized as follows: In the following section~\ref{sec:toymodel}, we introduce our toy quantum gravity model and discuss the expected UV behavior based on naive superficial degree of divergence. In section~\ref{sec:zero-momentum}, we will look at one- and two-loop quantum integrals at zero external momenta to assess whether the divergence structure confirms to the superficial degree of divergence. In section~\ref{sec:external-momentum}, we will look at $1$- and $2$-loop integrals with non-zero external momenta to identify both the divergent structure as well as the large (external) momentum behavior which is crucial in determining whether the theory can be renormalized loop-by-loop. In section~\ref{sec:dressed}, we will discuss the challenges in the renormalization prescription in the exponential infinite-derivative model, and demonstrate how this could be naturally addressed when one starts to use the dressed propagator instead of the bare propagator. In section~\ref{sec:conclusions}, we will conclude by summarizing our results.

%%%%%%%%%%%%%%%%%%%%%%%%%%%%%%%%%%%%
\section{Quantum gravity {\it toy model}}
\numberwithin{equation}{section}
\label{sec:toymodel}
%%%%%%%%%%%%%%%%%%%%%%%%%%%%%%%%%%%%%
\subsection{Superficial degree of divergence}\label{SDD}
Throughout this paper we will be interested in metric fluctuations, $h_ {\mu \nu}$, around the Minkowski background:
\be
g_{\mu \nu} = \eta_{\mu \nu} + h_{\mu \nu}\,.
\label{minkowski}
\ee
Gravity being a gauge theory only contains kinetic terms, \ie terms containing derivatives. In the case of GR all the terms contain two derivatives. In momentum space this means that the propagators behave as $k^{-2}$, while each vertex also comes with a $k^2$ factor. This is the compensating feature discussed in the introduction and is a hallmark of gauge theories.  Further in four dimensions, each momentum loop provides a $k^4$ factor in a quantum loop integral. The superficial degree of divergence of a Feynman diagram in GR is therefore given by (see~\cite{DeWitt:2007mi,Stelle:1976gc}):
\be
D = 4 L - 2 I + 2 V\,,
\ee
where $L$ is the number of loops, $V$ is the number of vertices, and $I$ is the number of internal propagators.
Using the topological relation:
\be
L = 1 + I - V\,,
\label{topological}
\ee
we get
\be
\label{eq:GR}
D = 2 L + 2\,.
\ee
Thus, the superficial degree of divergence increases as the number of loops increases, which is why GR is said to be non-renormalizable.

For Stelle's $4$th-order theory~\cite{Stelle:1976gc}, the graviton propagator goes as $\sim k^{-4}$, while the vertices $\sim k^4$, leading to a constant degree of divergence formula
\begin{equation}
D = 4\,.
\end{equation}
In other words, the degree of divergence does not increase with loops which enabled Stelle to prove that such a theory is renormalizable. Unfortunately, such a theory also contains a Weyl ghost which makes the theory non-unitary. As explained before, we will follow a different approach where we will introduce an infinite series of higher-derivative operators in a way that doesn't introduce any new states, ghosts or otherwise. We will see that the divergence counting will also be different as it will be based on the exponents rather than the degree of the polynomial momentum dependences.
%%%%%%%%%%%%%%%%%%%%%%%%%%%%%%%%%%%
\subsection{Infinite-derivative gravitational action}\label{intro-1}
The ``simplest'' infinite-derivative action that can modify the propagator of the graviton  without introducing any new states  is of the form~\cite{Biswas:2011ar,Biswas:2013kla}
\be
\label{action}
S = S_{EH} + S_{Q}\,,
\ee
where $S_{EH}$ is the Einstein-Hilbert action,
\be
\label{eq:EH}
\int d ^ 4 x \, \sqrt{-g} \, \frac{R}{2}\,,
\ee
and $S_{Q}$ is given by~\footnote{Around Minkowski space or in any maximally symmetric background it can be shown that ${\cal F}_3$ is redundant,
see~\cite{Biswas:2011ar,Biswas:2013kla}.}
\be
\label{nlaction}
S_{Q} = \int d ^ 4 x \, \sqrt{-g} \LT R \cF _ {1} (\Box) R + R _ {\mu \nu} \cF _ {2} (\Box) R ^ {\mu \nu} + R _ {\mu \nu \lambda \sigma} \cF _ {3} (\Box) R ^ {\mu \nu \lambda \sigma}\RT\,,
\ee
where the $\cF_i$'s are analytic functions of $\Box$ (the covariant d'Alembertian operator):
\be
\cF_{i} (\Box) = \sum _ {n=0}^{\infty} f_{i_n} \Box^{n}\,,
\label{quadratic}
\ee
satisfying~\footnote{For other forms of infinite-derivative gravity theories which contain an additional scalar degree of freedom, see~\cite{Biswas:2013kla,BT}.}
\be
\label{eq:kuku}
2\cF_1+\cF_2+2\cF_3=0
\ee
and the constraint that the combination
\be
\label{eq:a}
a (\Box) = 1 - \frac{1}{2} \mathcal{F} _ {2} (\Box) \Box - 2 \mathcal{F} _ {3} (\Box) \Box
\ee
is an {\it entire function}, with no zeroes. In Eq.~\eqref{quadratic}, the $f_{i_n}$'s  are real coefficients. Eqs.~\eqref{action}-\eqref{eq:a} define the BGKM gravity models. The classical equations of motion have been
studied for the above action~\cite{Biswas:2013cha}, and shown to be free from black-hole type of singularities for ``small'' central masses. In this paper therefore, we take the next logical step of investigating the quantum UV behavior of these theories.

For BGKM-type models, the quadratic (in $h_{\mu\nu}$) or ``free'' part of the action simplifies considerably, and one obtains~\footnote{There is also a part of action for one of the scalar modes of the metric, but this is a ghost degree of freedom that is precisely required to cancel the time-like contributions of the spin-two field~\cite{peter}.}:
\be
S_{\mt{free}}=\frac{M_{p}^{2}}{2}\int d^4x\ h^{\perp\mu\nu}\Box a(\Box)\hp_{\mu\nu}\,,
\ee
where $\hp_{\mu\nu}$
is the transverse traceless spin $2$ graviton mode, satisfying:
\be
\n^{\mu}\hp_{\mu\nu}=g^{\mu\nu}\hp_{\mu\nu}=0\,.
\ee
This leads to the propagator~\cite{Biswas:2011ar,Biswas:2013kla},
\be
\Pi(k^2) = - \frac{i}{k^2a(-k^2)}\LF {\cal P}^2 - \frac{1}{2} {\cal P}_s ^0  \RF=\frac{1}{a(-k^2)} \Pi_{GR}\,,
\ee
for the physical degrees of freedom for the graviton (see~\cite{Biswas:2013kla,peter} for the definitions of the spin projector operators ${\cal P}^2$ and ${\cal P}_s ^0$).

Ideally, we should now compute the interaction terms for our non-local gravity theory and then use it to compute the Feynman diagrams. This however turns out to be an extremely challenging task for several reasons: Gravitational theories are all order theories and therefore contain interactions of all orders in $ h_{\mu\nu}$, and computing all these interactions is well beyond the scope of the current paper. While one can argue that all the terms which are higher order in fields have additional Planck suppressions~\footnote{The easiest way to see this is to redefine $\hp_{\mu\nu}\ra M_p\hp_{\mu\nu}$, which is anyway required to make the free terms, Eq.~\eqref{free},  canonical. Then, each additional field comes with an additional Planck suppressed factor making them sub-leading to the cubic interactions in the low energy limit.}, and that therefore the most relevant piece in the low energy approximation comes from the cubic terms, unfortunately even computing the complete cubic interactions for an action such as Eq.~\eqref{nlaction} is challenging. Moreover, the expressions are rather complicated making further progress in evaluating Feynman diagrams very difficult.

Therefore, rather than taking on this arduous task, in this paper we wish to  understand whether non-localization can at all help to tame the UV divergences in gravity theories, given the compensating nature of the exponential suppressions and enhancements present in the propagators and vertices. To avoid getting muddled in complex algebra we will try to understand the physics in a simple scalar {\it toy model} which we will arrive at using symmetry principles that helps us to retain some of the crucial properties of the full gravitational theories, see Appendix~\ref{sec:BRST} for details. For a comparative study and future reference we have however included some details about the gravitational action in  Appendix~\ref{sec:prototype},
where we have calculated some of the prototype cubic interaction terms that one obtains from Eq.~\eqref{action}~\footnote{One way to obtain a toy scalar field model which mimics the gravitational Lagrangian is to substitute a conformally flat metric $h_{\mu \nu} =\Omega^2(x) \, \eta_{\mu \nu}$ in~\eqref{nlaction}. The scalar field action that can be obtained this way is similar to the toy model we will consider, but its  kinetic term has the wrong sign  because it basically corresponds to the unphysical ghost degree in $P_s^0$, and also contains additional terms involving double sums that again makes the model technically more challenging to deal with. We therefore adopt a cleaner strategy based on symmetries to obtain the scalar toy model.}.

%%%%%%%%%%%%%%%%%%%%%%%%%%%%

\subsection{Motivating scalar {\it toy model} of quantum gravity from symmetries}
\label{sec:scalartoymodel}
It is well known that the field equations of GR exhibit a global scaling symmetry,
\be
g_{\mu\nu}\ra \la g_{\mu\nu}\,.
\label{scaling}
\ee
%Any quadratic curvature action of the form Eq.~(\ref{nlaction}) itself is also invariant under Eq.~(\ref{scaling}).
When we expand the metric around the
Minkowski vacuum, Eq.(\ref{minkowski}), the scaling symmetry translates to a symmetry for $h_{\mu\nu}$, whose infinitesimal version is given by
\be
h_{\mu\nu} \to (1 + \epsilon) h_{\mu\nu} + \epsilon\eta_{\mu\nu}\,.
\ee
While we do not expect the scaling symmetry to be an unbroken fundamental symmetry of nature, the symmetry serves a rather useful purpose for us.
It relates the free and interaction terms just like gauge symmetry does. Thus, we are going to use this combination of shift and scaling symmetry
\be
\phi \to (1 + \epsilon) \phi + \epsilon\,,
\label{scale-shift}
\ee
to arrive at a scalar {\it toy model} whose propagator and vertices preserve the compensating nature found in the full BGKM gravity.
Inspired by the discussion in the previous section, we will now consider a scalar {\it toy model} with a string field theory type free action:
\be
\label{free}
S_{\mt{free}} = \frac{1}{2}\int d^4 x \, \LF  \phi \Box a(\Box) \phi\RF\,,
\ee
where for the purpose of this paper, we are going to choose~\cite{Biswas:2011ar,Biswas:2013kla}:
\be
a (\Box) = e^{- \Box / M ^ {2}}\,,
\ee
where $M$ is the mass scale at which the non-local modifications become important. In general, one is free to choose any entire function,
while keeping in mind that $a(k^2)\rightarrow 1$ for the IR momentum, $k\rightarrow 0$, in order to recover the propagator of the usual GR.  Note that the
sign of $a(\Box)$ is also crucial in order to recover the correct Newtonian potential as shown in Ref.~\cite{Biswas:2011ar,Biswas:2013kla}~\footnote{If we had chosen $a(\Box)$ to be
\be
a(\Box) = e ^ {\Box/M^2}\,,
\ee
where $M^2>0$, then we can perform the loop integrals for $a(\Box) = e ^{-\Box/M^2}$, assuming $M^2>0$, and then analytically continue to $M^2<0$.  In this way the Newtonian potentials $\Phi(r)$ \& $\Psi(r)$ would be given by
\begin{align}
\Psi(r)=\Phi(r)
 = \frac{2 i m \pi ^ 2}{M _ {p} ^ {2} r} \, \mathbf{Erfi} \LF \frac{M r}{2} \RF \where \mathbf{Erfi} (z) = \frac{\mathbf{Erf}(i z)}{i}
\end{align}
is the imaginary error function and admits real values for real $z$. Clearly in this case, the Newtonian potential  is purely imaginary indicating an unphysical theory. See also~\cite{Modesto:2014eta} for a discussion of Newtonian singularities in higher-derivative gravity models.}.

The symmetry, Eq.~(\ref{scale-shift}), then uniquely fixes the cubic interaction term, see Appendix~\ref{sec:A},
\be
\label{int}
S_{\mt{int}} = \frac{1}{M_p} \int d ^ 4 x \, \LF \frac{1}{4} \phi \partial _ {\mu} \phi \partial ^ {\mu} \phi + \frac{1}{4} \phi \Box \phi a(\Box) \phi - \frac{1}{4} \phi \partial _ {\mu} \phi a(\Box)  \partial ^ {\mu} \phi \RF\,,
\ee
up to integrations by parts. Our  {\it toy model} action is then given by:
\be
\label{eq:action}
S_{\mt{scalar}} = S _ {\mt{free}} + S _ {\mt{int}}\,.
\ee
It is now time to revisit the superficial degree of divergence for this {\it toy model}.

Since an exponential suppression always dominates over a polynomial growth, the naive expectation is that as long as the exponentials come with a negative power, the integrals should converge. Thus, rather than computing the power of polynomial divergence in momentum, we are really interested in calculating the pre-factor in the exponent, and this radically changes the counting of the superficial degree of divergence.

Since every propagator comes with an exponential suppression, see Eq.~(\ref{free}), while every vertex comes with an exponential enhancement, Eq.~(\ref{int}), the superficial degree of divergence counting in exponents is given by
\be
E = V - I\,.
\ee
By using the topological relation, Eq.~\eqref{topological}, we obtain:
\be
\label{eq:E}
E = 1 - L\,.
\ee
Thus, except for the $L = 1$ loop, $E < 0$, and the corresponding loop amplitudes are superficially convergent. In Appendix \ref{sec:BRST}, we discuss the analogous calculation in the complete BRST-invariant quantum infinite-derivative gravitational action. The conclusion is exactly the same. The rest of the paper is devoted to investigating whether the naive expectation about the convergence properties of the Feynman diagrams hold up in explicit calculations involving quantum loop calculations.

%%%%%%%%%%%%%%%%%%%%%%%%%%%
\section{Divergence structure with zero external momenta}\label{sec:zero-momentum}
\numberwithin{equation}{section}
%%%%%%%%%%%%%%%%%%%%%%%%%%%
\subsection{Feynman rules}
%%%%%%%%%%%%%%%%%%%%%%%%%%
All the Feynman rules and Feynman integral computations in this paper are carried out in Euclidean space after analytic continuation ($k_{0} \to i k _ {0}$ \& $k^2 \to k_{E}^2$ using the mostly plus metric signature; we shall drop the $E$ subscript for notational simplicity).
The final results we obtain can then be analytically continued back to Minkowski space as desired.

The Feynman rules for our action Eqs.~\eqref{free} and \eqref{int} can be derived rather straightforwardly. The propagator in momentum space is then given by
\be
\Pi (k ^ 2)= \frac{- i}{k^2 e ^ {\kb ^ 2}}\,,
\ee
where barred $4$-momentum vectors from now on will denote the  momentum divided by the mass scale $M$. The vertex factor for three incoming momenta $k_{1},~k_{2},~k_{3}$ satisfying the conservation law:
\be
k _ {1} + k _ {2} + k _ {3} = 0\,,
\label{conservation}
\ee
is given by
\be
\label{eq:V}
\frac{1}{M_{p}}V (k _ {1}, k _ {2}, k _ {3}) = \frac{i}{M_p} C(k_1,k_2,k_3) \LT 1 -  e ^ {\kb _ {1} ^ {2}} -  e ^ {\kb _ {2} ^ {2}} - e ^ {\kb _ {3} ^ {2}}\RT\,,
\ee
where
\be
C (k_1,k_2,k_3)= \frac{1}{4} \LF k _ {1} ^ {2} + k _ {2} ^ {2} + k _ {3} ^ {2} \RF\,.
\ee
Let us briefly explain how we obtain the vertex factor.
The first term originates from the term, $ \4 \phi \partial _ {\mu} \phi \partial ^ {\mu} \phi$, which using Eq.~\eqref{conservation} in the momentum space, reads
\ba
- \frac{i}{2} (k _ {1} \cdot k _ {2} + k _ {2} \cdot k _ {3} + k _ {3} \cdot k _ {1})
=  \frac{i}{4} \left(k _ {1} ^ {2} + k _ {2} ^ {2} + k _ {3} ^ {2} \right) \,.
\ea
The second term comes from the terms, $\frac{1}{4} \phi \Box \phi a(\Box) \phi$, and $- \frac{1}{4} \phi \partial _ {\mu} \phi a(\Box)  \partial ^ {\mu} \phi$. In the
momentum space, again using Eq.~\eqref{conservation}, we get
\be
\frac{i}{4} \left(k _ {3} \cdot k _ {1} + k _ {1} \cdot k _ {2} - k _ {3} ^ {2} - k _ {2} ^ {2} \right) e ^ {\kb _ {1} ^ {2}}  = - \frac{i}{4} \left(k _ {1} ^ {2} + k _ {2} ^ {2} + k _ {3} ^ {2} \right) e ^ {\kb _ {1} ^ {2}} \,.
\ee
 The third and the fourth terms in Eq.~\eqref{eq:V} arise in an identical fashion.

For future convenience, let us consider the special case when one of the momenta is zero. For instance,
choosing $k _ {3} = 0$, we obtain $k _ {1} = - k _ {2} = k$, which then gives us
\be
V(k)\equiv V (k, -k, 0) = - i k ^ {2} e ^ {\kb ^ {2}}\,.
\ee
We will also often encounter the square of the vertex factor, which is  given by:
\begin{align}
\label{Vsquare}
V^ {2} \LF k _ {1}, k _ {2}, k _ {3} \RF
& = i ^ {2} C ^ {2}(k_1,k_2,k_3)\LT1 - 2  e ^ {\kb _ {1} ^ {2}} - 2  e ^ {\kb _ {2} ^ {2}} - 2 e ^ {\kb _ {3} ^ {2}}+ 2  e ^ {\kb _ {1} ^ {2}} e ^ {\kb _ {2} ^ {2}} \Rd \non
& + \Ld 2  e ^ {\kb _ {2} ^ {2}} e ^ {\kb _ {3} ^ {2}} + 2 e ^ {\kb _ {3} ^ {2}} e ^ {\kb _ {1} ^ {2}} + e ^ {2 \kb _ {1} ^ {2}} +  e ^ {2 \kb _ {2} ^ {2}} +  e ^ {2 \kb _ {3} ^ {2}}\RT\,.
\end{align}

%%%%%%%%%%%%%%%%%%%%%%%%%%%%%%%%%%%%%%%%%%%%%%%%%%%%%%%%%%

\subsection{$1$-loop, $2$-point function with zero external momenta}

\begin{figure}[t]
\centering
\includegraphics[width=.35\textwidth]{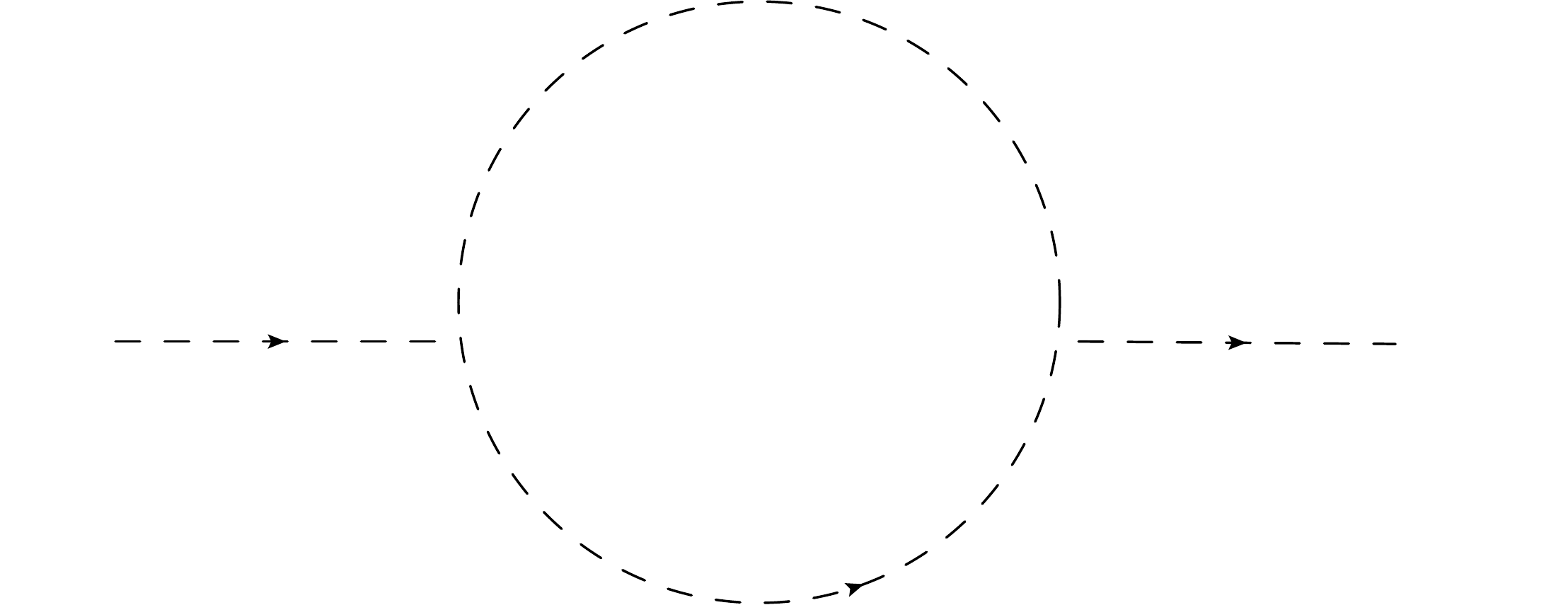}\hspace{3cm}
\includegraphics[width=.35\textwidth]{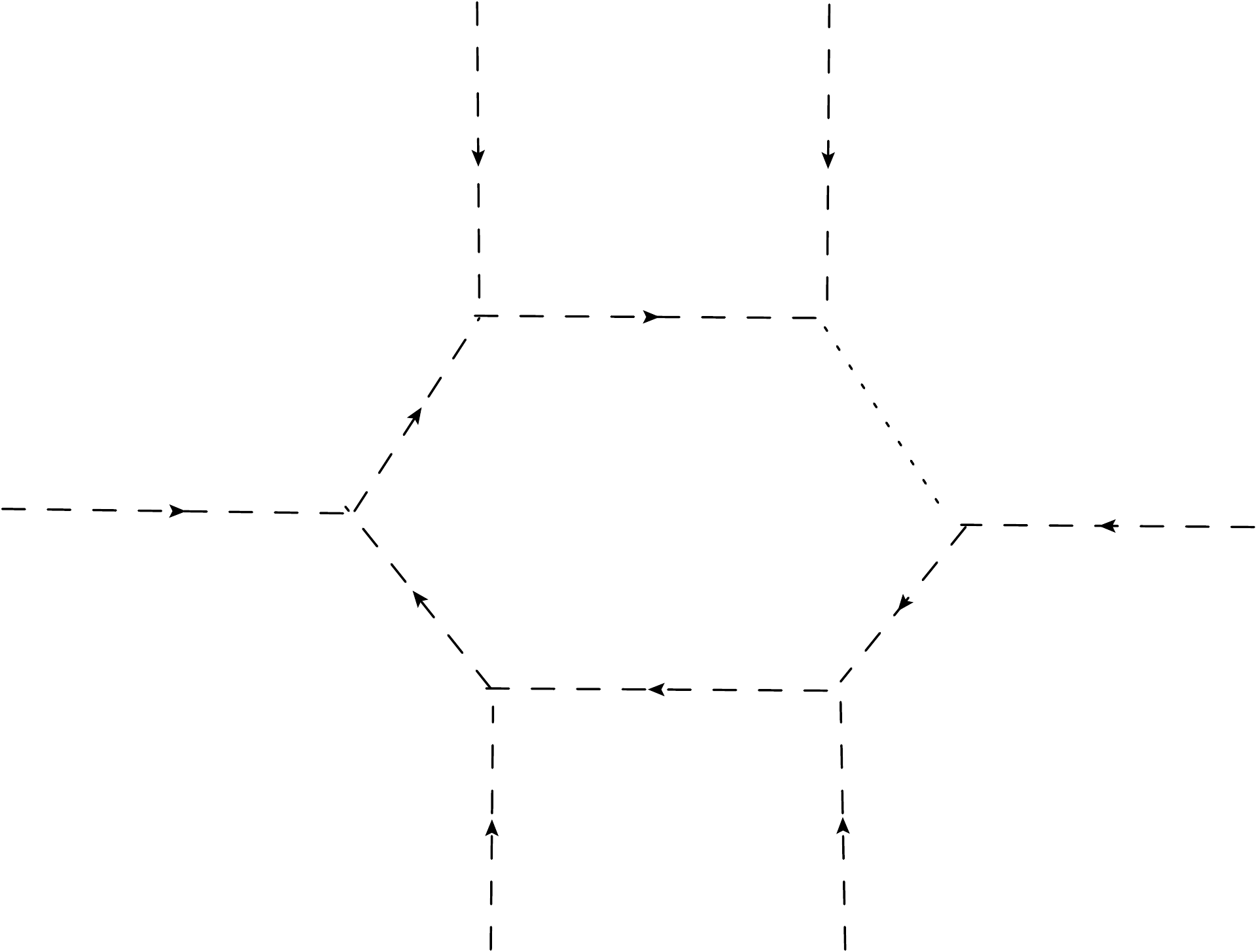}
\caption{\label{fig:1-loop} {\small Left: $1$-loop, $2$-point diagram $\Ga_2$. Right: The $1$-loop, $N$-point diagram $\Ga_{N}$. The dots indicate an arbitrary number of (bare) vertices and (bare) propagators for the scalar field.}}
\end{figure}

Let us start with the $1$-loop $2$-point function. There is only one Feynman diagram as depicted in Fig. \ref{fig:1-loop} (left). According to the Feynman rules, we have~\footnote{The mass correction is naively given by $\delta m^2 = i \Ga_2$ (which is negative). When we derive the dressed propagator in section~\ref{sec:dressed}, we will get the exact mass correction.},
\be
\Ga_2 = \frac{i}{2 M _ {p} ^ {2}} \int \frac{\mathrm{d} ^ 4 k}{(2 \pi) ^ 4} \, \frac{V ^ {2} (k)}{i ^ {2} k ^ 4 e ^ {2 \kb ^ {2}}}\,.
\ee
Note that we are working in an Euclidean space and that the symmetry factor is $2$. The angular integrations can be performed trivially, see Appendix~\ref{sec:D1} for details, as the integrand only depends on the norm of the external momentum, leaving us with
\be
\Ga_2 = \frac{i}{2 M _ {p} ^ {2}} \frac{4 \pi}{(2 \pi) ^ 4} \int _ {0} ^ {\Lambda} \mathrm{d} k \, \frac{\pi  k^3}{2}\,.
\ee
Integrating with respect to $k$ from $0$ to $\Lambda$, where $\Lambda$ is a hard cutoff, we obtain:
\be
\Ga_2 =\frac{i \Lambda ^4}{64 M _ {p} ^ {2} \pi ^ 2}\,.
\ee
We see that the integral goes like $\int \mathrm{d} ^ 4 k$, and is therefore sensitive to the UV cut-off.
This result is in complete accordance with the analysis of superficial degree of divergence according to which at
$1$-loop level the exponential non-locality does not affect the integrals. The divergence structure is exactly the same as that of GR at $1$-loop.
%%%%%%%%%%%%%%%%%%%%%%%%%%%%%%%%%%%%%%%%%%%%%%%%%%%%%%%%%%%

\subsection{$N$-point function with zero external momenta}
An interesting fact for gravitational theories is that the superficial degree of divergence does not depend on the number of external vertices. This is true both in
GR, see Eq.~(\ref{eq:GR}), as well as in BGKM gravity and in infinite-derivative scalar field theory, see Eq.~(\ref{eq:E}). Let us then calculate the $N$-point function at one loop, see Fig.~\ref{fig:1-loop} (right). As one can see, the $N$-point diagram is not particularly different from the $2$-point diagram; it is an $N$-polygon with $N$ vertices and $N$ edges. Thus, instead of a square of the propagator and vertex factor, one now has $N$ powers of them:
\be
\Ga_N = \frac{i}{M _ {p} ^ {N}} \int \frac{\mathrm{d} ^ 4 k}{(2 \pi) ^ 4} \, \frac{V ^ {N} (k)}{i ^ {N} k ^ {2 N} e ^ {N \kb ^ {2}}}= (-1) ^ {N} \frac{i \Lambda ^4}{32 M _ {p} ^ {N} \pi ^ 2}\,,
\ee
where again $\Lambda$ is the hard cutoff. As expected, its divergence is the same as that of the $2$-point function  precisely as predicted by the divergence power-counting. The above diagrams are also known as ring diagrams, and they contribute to the {\it effective potential}. The symmetry factor is $2 N$~\footnote{It should be noted that the symmetry factor is equal to $2N$ when $1$PI corrections to the effective potential are considered (the external points are not fixed in that case). When computing a Green's function, the symmetry factor is equal to $2$ for $N=1,2$ and to $1$ for $N>2$.}, and summing all the $1$-loop diagrams, one obtains the one-loop contribution to the effective potential,
\be
V_{\mathrm{eff}} ^ {(1)} (\phi) = i \sum_{N=1}^{\infty} \Ga_{N} \phi^N = \sum_{N=1}^{\infty} \frac{(-1) ^ {N +1}}{2N} \frac{\Lambda ^4}{32 M _ {p} ^ {N} \pi ^ 2} \phi^N = \frac{\La^4}{64 \pi^2} \ln \LF 1 + \frac{\phi}{M_p} \RF\,,
\ee
as is typical, see also Refs.~\cite{Bluhm,Reddy,CW} for similar computations. In a theory of gravity, we of course, do not expect to find such an effective potential as that would violate general covariance, diagrams coming from different order interactions must cancel the contributions. We obtain these terms in our {\it toy model} since the scaling symmetry is only a symmetry of the field equations and not the entire action, and therefore it is expected to be broken at the quantum level.

The prescription that was used in Ref.~\cite{Reddy} to eliminate these divergent terms while preserving the pole mass is to simply add an opposing counter term. This is also the prescription that is followed in standard field theoretic calculations (renormalization conditions), and we will adopt the same convention as we move on to higher loop diagrams.

Our calculations  corroborated the expected divergence structure, Eq.~\eqref{eq:E}, in infinite-derivative theories, or any covariant theory of gravity for that matter. To prove renormalizability, the real challenge will be to demonstrate that once these $1$-loop divergences (subdivergences) are eliminated by counterterms in higher loop subdiagrams, the remaining loop integrals yield finite result. At the least this means that the higher than 1-loop diagrams cannot diverge more than the bare vertex. This is what we now want to check in the remaining sections.

%$\frac{i \La^4}{32 M _ {p} \pi^2} \LF \frac{1}{M_{p}+1} - \frac{1}{2 M _ %{p}} \RF$
%%%%%%%%%%%%%%%%%%%%%%%%%%%%%%%%%%%
\subsection{$2$-point function at $2$-loop order}
\begin{figure}[t]
\centering % \begin{center}/\end{center} takes some additional vertical space
\includegraphics[width=.35\textwidth]{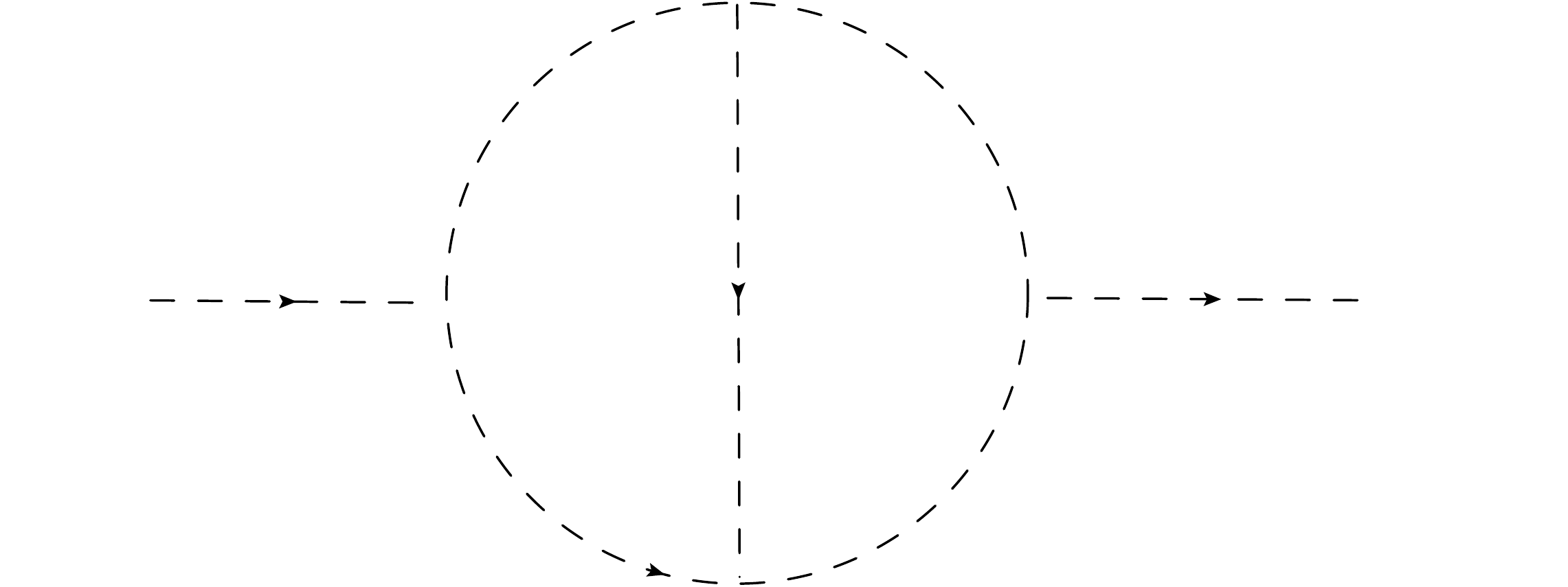}
\includegraphics[width=.40\textwidth]{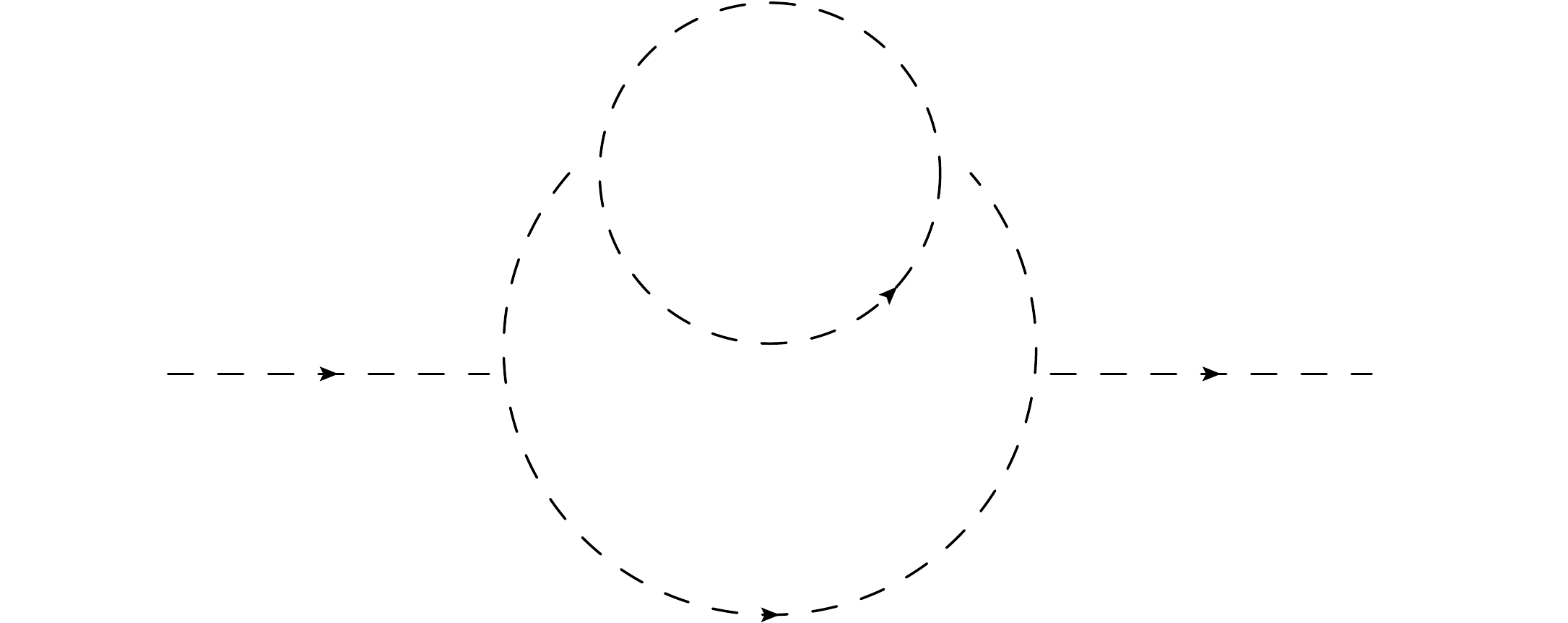}
 %"%\includegraphics" is very powerful; the graphicx package is already loaded
\caption{\label{fig:2-loop} {\small Left: The $2$-loop, $2$-point diagram $\Ga_{2,2a}$. Right: The $2$-loop, $2$-point diagram $\Ga_{2,2b}$.}}
\end{figure}

%%%%%%%%%%%%%%%%%%%%%%%%%%%%%%%%%%
\subsubsection{General structure}
We wish to now investigate the second feature of the divergence formula, Eq.~(\ref{eq:E}), namely, that for higher than $1$-loop no new divergences should emerge. Since there are always subdivergent $1$-loop graphs within a $2$-loop diagram, we do not expect in general finite results, but what we wish to find here is that the $2$-loop graph should have the same divergence behavior as that of the $1$-loop counterpart. In other words, they should diverge at most as $\La^4$. This result will be in contrast with the GR case, where the $2$-loop diagrams diverge as $\La^6$.

There are two Feynman diagrams as depicted in Fig.~\ref{fig:2-loop}. The Feynman diagram with zero external momenta in
Fig.~\ref{fig:2-loop} (left) is given by
\be
\Ga_{2,2a} = \frac{i ^ {2}}{2 i ^ {5} M _ {p} ^ {4}} \int \frac{\mathrm{d} ^ 4 k _ {1}}{(2 \pi) ^ 4} \frac{\mathrm{d} ^ 4 k _ {2}}{(2 \pi) ^ 4} \, \frac{V (k _ {1}) V (k_ {2}) V ^ {2} (k _ {1}, k _ {2},k _ {3})}{k _ {3} ^ {2} k _ {2} ^ {4} k _ {1} ^ {4} e ^ {\kb _ {3} ^ {2}} e ^ {2 \kb _ {2} ^ {2}} e ^ {2 \kb _ {1} ^ {2}}}\,,
\ee
where $k _ {3} = - k _ {1} - k _ {2}$, and the expression is symmetric in $k _ {1}$ and $k _ {2}$. The numerator contains a sum of different exponents, so that the overall integral can be written in the form
\be
\label{eq:44}
\Ga_{2,2a} = \frac{i}{2 M _ {p} ^ {4}} \int \frac{\mathrm{d} ^ 4 k _ {1}}{(2 \pi) ^ 4} \frac{\mathrm{d} ^ 4 k _ {2}}{(2 \pi) ^ 4} \, \frac{C ^ {2}}{k _ {1} ^ {2} k _ {2} ^ {2} k _ {3} ^ {2}}\sum_i \lambda _ {i} \exp[E_i(k_1,k_2)]\,,
\ee
where  $E_i$'s are quadratic polynomials of $k_1,~k_2$ and $\lambda _ {i}$ are constants taking on the values $-2$, $-1$, $+1$, $+2$. Firstly, let us note that one can always find linear combinations of $k_1,~k_2$, lets call them $q_1,~q_2$, such that $E_i$ is diagonal:
\be
E_i=a_1q_1^2+a_2q_2^2\,.
\ee
Now, depending upon the value of the $a_i$'s one can classify the terms in three groups:

\begin{enumerate}

\item[(I)] If both $a_1,~a_2< 0$, both the momentum integrals can be performed to provide a finite answer.

\item[(II)] If both $a_i$'s are nonzero, but one of them is positive, then  one can obtain the integrals by suitably analytically
continuing results from the $a_i<0$ to $a_i>0$ region.
%~\footnote{\bf comment of wick rotation.}.

\item[(III)] Finally, there are cases when one of the $a_i$'s is zero. We expect that this represents the divergent contribution from the $1$-loop subdiagram embedded within the $2$-loop graph. We shall check whether this provides a $\La^4$ divergence, or a $\La^6$ as in usual GR.

\end{enumerate}
%%%%%%%%%%%%%%%%
\subsubsection{Convergent  groups (I)  terms}
Let us first look at the  Group (I) terms. The overall exponential factors for the group (I) type terms are given by
\ba
e ^ {- \kb _ {1} ^ {2}} e ^ {- \kb _ {2} ^ {2}} e ^ {- \kb _ {3} ^ {2}}, ~~~~~~~e ^ {- \kb _ {2} ^ {2}} e ^ {- \kb _ {1} ^ {2}}, ~~~~~~~e ^ {- \kb _ {1} ^ {2}}  e ^ {- \kb _ {3} ^ {2}}, ~~~~~~~
e ^ {- \kb _ {2} ^ {2}} e ^ {- \kb _ {3} ^ {2}}.
\label{group-one}
\ea
The first integrand evaluates to
\be
\frac{3i M^6 \log \left({4}/{3}\right)}{4096 M _ {p} ^ {4} \pi ^4}\,,
\ee
%See the Mathematica notebook file 25-May-2014.nb.
while all the other three gives us the same contribution:
\be
-\frac{iM^6 (3+\log (4))}{2048 M _ {p} ^ {4} \pi ^4}\,.
\ee
%See the Mathematica notebook file 24-May-2014.nb.
Thus, all together we have
\be
\Ga_{2,2,\mt{i}} =\frac{3i M^6}{2048 M _ {p} ^ {4} \pi ^4}\LT\frac{1}{2} \log \left(\frac{4}{3}\right) - (3+\log 4)\RT\,.
\ee
%%%%%%%%%%%%%%%%%%%%%%%%%%%%%%%%%%%%%%%%%%%%%%%%%%%%%%%%%
\subsubsection{Group (II) \& (III) terms \& the divergence structure}
Next let us look at the integrals originating from the last three terms in Eq.~(\ref{Vsquare}). With overall exponents
\ba
e ^ {\kb _ {1} ^ {2}} e ^ {- \kb _ {2} ^ {2}} e ^ {- \kb _ {3} ^ {2}}~, ~~~~~~~e ^ {\kb _ {2} ^ {2}} e ^ {- \kb _ {1} ^ {2}}e ^ {- \kb _ {3} ^ {2}}~, ~~~~~~~
e ^ {\kb _ {3} ^ {2}}  e ^ {- \kb _ {1} ^ {2}} e ^ {- \kb _ {2} ^ {2}}\,,
\label{group-two}
\ea
these form the group (II) set with one eigenvalue positive and one negative. These integrals can also be evaluated by employing suitable analytic continuation methods, please see Appendix~\ref{sec:D2} for details. 

Again, all the terms contribute equally, and we get
\be
\label{eq:imaginary}
\Ga_{2,2,\mt{ii}} = \frac{3iM^4}{4096 \pi^{4} M_{p}^{4}} \left(M^2 \left(\log (4)-8\right)-4 \Lambda ^2 \right).
\ee

We are left to tackle the group (III) terms  originating from the fifth, sixth and seventh terms in Eq.~(\ref{Vsquare}), whose exponential contributions coming from  the vertices are given by
\ba
e ^ {\kb _ {1} ^ {2}} e ^ {2 \kb _ {2} ^ {2}} e ^ {\kb _ {3} ^ {2}}, ~~~~~~~e ^ {\kb _ {2} ^ {2}} e ^ {2 \kb _ {1} ^ {2}} e ^ {\kb _ {3} ^ {2}}, ~~~~~~e ^ {2 \kb _ {1} ^ {2}} e ^ {2 \kb _ {2} ^ {2}}\,.
\label{group-three}
\ea
Since, the exponential contribution of the propagators is given by:
\be
e ^ {- 2 \kb _ {1} ^ {2}} e ^ {- 2 \kb _ {2} ^ {2}} e ^ {- \kb _ {3} ^ {2}}\,,
\ee
the overall exponents, $E_i$, for the three above cases are
\be
- \kb _ {1} ^ {2},~~~~ - \kb _ {2} ^ {2},~~~~ - \kb _ {3} ^ {2}\,.
\ee

Clearly, there is no exponential damping along the directions orthogonal to $k_1$, $k_2$ and $k_3$ respectively.
Accordingly, while one of the momentum integrals is convergent due to the presence of the exponential, the other one can only be computed using a hard cutoff. The result is identical for all the three diagrams, and one obtains
\begin{align}
\Ga_{2,2,\mt{iii}} &= \frac{i M ^ 2}{4096 M _ {p} ^ {4} \pi ^ 4} \left[\vphantom{\frac{\Lambda ^6 \text{Ei}\left(-\frac{\Lambda ^2}{M^2}\right)}{M ^ 2}}  2  M^4 \left(6 \log \left(\frac{\Lambda}{M} \right)+5 e^{-\frac{\Lambda ^2}{M^2}}-3 Ei \left(- \frac{\Lambda ^2}{M^2}\right)+3 \gamma -5\right) \right. \non
&+ \left. \Lambda ^2 M^2 \left(e^{-\frac{\Lambda ^2}{M^2}}+15\right) + \Lambda ^4 \left(6-e^{-\frac{\Lambda ^2}{M^2}}\right) -\frac{\Lambda ^6 Ei\left(-\frac{\Lambda ^2}{M^2}\right)}{M ^ 2}\right]\,,
\end{align}
where
\be
Ei (z) \equiv - \dashint _ {- z} ^ {\infty} \mathrm{d} t \, \frac{e ^ {-t}}{t}
\ee
is the exponential-integral function, see Ref.~\cite{Hassani}, and has a branch cut discontinuity in the complex $z$-plane running from $0$ to $\infty$. The sign $\dashint$ indicates that the principal value of the integral is taken. We note that, for large negative $z$, the $Ei$ function falls off as a Gaussian and therefore can be ignored in the $\La\ra \infty$ limit. Therefore, the surviving divergent pieces read
\be
\Ga_{2,2,\mt{iii}} = \frac{iM ^ 2}{4096 M _ {p} ^ {4} \pi ^ 4} \left[
12  M^4 \log \left(\frac{\Lambda}{M} \right)+ 15\Lambda ^2 M^2  + 6\Lambda ^4 +2  M^4 \left(3 \gamma -5\right) \RT\,.
\ee
Firstly, we see that the divergence is indeed  $\propto \La^4$, as the superficial divergence argument suggested, and does not grow as $\La^6$ that one would find in GR.

We note that all the results obtained in this section have been divided by a symmetry factor $2$ for the diagram. Summing all the integrals, we obtain our final result:
\be
\label{eq:6666}
\Ga_{2,2a} =\frac{iM ^ 2}{4096 M _ {p} ^ {4} \pi ^ 4} \LT \vphantom{\frac{\Lambda ^6 \text{Ei}\left(-\frac{\Lambda ^2}{M^2}\right)}{M ^ 2}} M ^ {4} \left(12 \log \left(\frac{\Lambda}{M}\right)-52+2\LF3 \ga-5 \RF- 3 \log (3)\right)  + 3 \Lambda ^2 M^2 +6\Lambda ^4 \RT.
\ee
To reiterate, $\Ga_{2,1}\sim\Ga_{2,2}\sim \La^4$, as the counting of  superficial degree of divergence would suggest. While we have not explicitly
calculated higher than two loop graphs, we would expect the same pattern to continue to hold, \ie, we do not expect larger than quartic
divergence in any loop order.

%%%%%%%%%%%%%%%%%%%%%%%%%%%%%%%%%%%%%
\subsubsection{The other $2$-loop diagram}
After setting the external momenta equal to zero, the Feynman diagram in Fig.~\ref{fig:2-loop} (right) becomes
\be
\Ga_{2,2b} = \frac{i ^ {2}}{2 i ^ {5} M _ {p} ^ {4}} \int \frac{\mathrm{d} ^ 4 k _ {1}}{(2 \pi) ^ 4} \frac{\mathrm{d} ^ 4 k _ {2}}{(2 \pi) ^ 4} \, \frac{V ^ {2} (k _ {1}) V ^ {2} (k _ {1}, -\frac{k _ {1}}{2} + k _ {2},-\frac{k _ {1}}{2}- k _ {2})}{k _ {1} ^ {6} (\frac{k _ {1}}{2} + k _ {2}) ^ {2} (\frac{k _ {1}}{2} - k _ {2}) ^ {2} e ^ {3 \kb _ {1} ^ {2}} e ^ {\LF\frac{\kb _ {1}}{2} + \kb _ {2}\RF ^ {2}} e ^ {\LF\frac{\kb _ {1}}{2} - \kb _ {2}\RF ^ {2}}}\,,
\ee
where we have assumed symmetrical routing of momenta and the symmetry factor of the diagram is $2$. Again, the numerator contains a sum of different exponents, so that the overall integral can be written in the form
\be
\label{eq:888}
\Ga_{2,2b} = \frac{i}{2 M _ {p} ^ {4}} \int \frac{\mathrm{d} ^ 4 k _ {1}}{(2 \pi) ^ 4} \frac{\mathrm{d} ^ 4 k _ {2}}{(2 \pi) ^ 4} \, \frac{D ^ {2}}{k _ {1} ^ {2} (\frac{k _ {1}}{2} + k _ {2}) ^ {2} (\frac{k _ {1}}{2} - k _ {2}) ^ {2}}\sum_i \mu _ {i} \exp[F_i(k_1,k_2)]\,,
\ee
where  $F_i$'s are quadratic polynomials of $k_1,~k_2$, and $\mu _ {i}$ are constants which take on the values $-2$, $-1$, $+1$, $+2$, similar to the first 2-loop diagram. Also,
\be
D = \frac{1}{4} \left(k _ {1} ^ {2} + \left(\frac{k _ {1}}{2} + k _ {2} \right) ^ {2} + \left(\frac{k _ {1}}{2} - k _ {2} \right) ^ {2} \right)\,.
\ee
If we change variables $k_1 \to k _ {1} ^ {'}$, ~~$- \frac{k _ {1}}{2} - k _ {2} \to k _ {2} ^ {'}$ (or, equivalently, $k _ {1} \to k _ {1} ^ {'}$ and  $- \frac{k _ {1}}{2} + k_{2} \to k _ {2} ^ {'}$) in $\Ga _ {2,1}$, we get $\Ga _ {2,2}$, since the Jacobian is $1$; \ie, Eq.~\eqref{eq:888} is exactly equivalent to Eq.~\eqref{eq:44}:
\be
\Ga_{2,2b} =\Ga_{2,2a}\,.
\ee
 Hence, the results for both the 2-loop diagrams are exactly the same.
To reiterate, $\Ga_{2,1}\sim\Ga_{2,2}\sim \La^4$, which would seem to corroborate the counting of superficial degree of divergence \eqref{eq:E}.
%%%%%%%%%%%%%%%%%%%%%%%%%%%%%%%%%%%%%%%%%%%%%%%%%%%%%%%%%%
\section{External momentum dependence and renormalizability}\label{sec:external-momentum}
\numberwithin{equation}{section}
%%%%%%%%%%%%%%%%%%%%%%%%%%%%%%%%%%%%%%%%%%%%%%%%%%%%%%%%%%%
\subsection{Arbitrary loop diagrams}

The calculations in the earlier subsection supported our naive divergence counting argument in section~\ref{sec:toymodel}, which suggested that all $1$-loop diagrams should be divergent $\sim \La^4$, and that this divergence should not increase as we go to higher loops. While this agreement is encouraging, just the fact that the divergence doesn't increase at higher loops doesn't  guarantee renormalizability. To achieve renormalizability, one has to check, for instance,  that once the $1$-loop sub-divergences are removed from a higher loop diagram, the diagram becomes finite. This requires keeping track of the UV behavior of the external momenta while performing the various loops. Let us illustrate the point with a few examples.

Consider the $2$-loop diagram in Fig.~\ref{fig:2-loop}, left. This contains a sub-divergent $3$-point, $1$-loop diagram. If we had found a prescription to make the $1$-loop diagram finite (for instance by adding appropriate counter terms as suggested in the previous section~\footnote{We will later see that the $3$-point function is actually finite once we introduce the dressed propagator.}), then the $2$-loop diagram should really be replaced by Fig.~\ref{2Lmodified} (left), where we now have a finite renormalized $3$-point function. We then have to perform a loop integral involving the renormalized $3$-point function:
%%%%%%%%%%%%%%%%%%%%%%%%%%%%%%%%%%%
\begin{figure}[t]
\centering
\includegraphics[width=.45\textwidth]{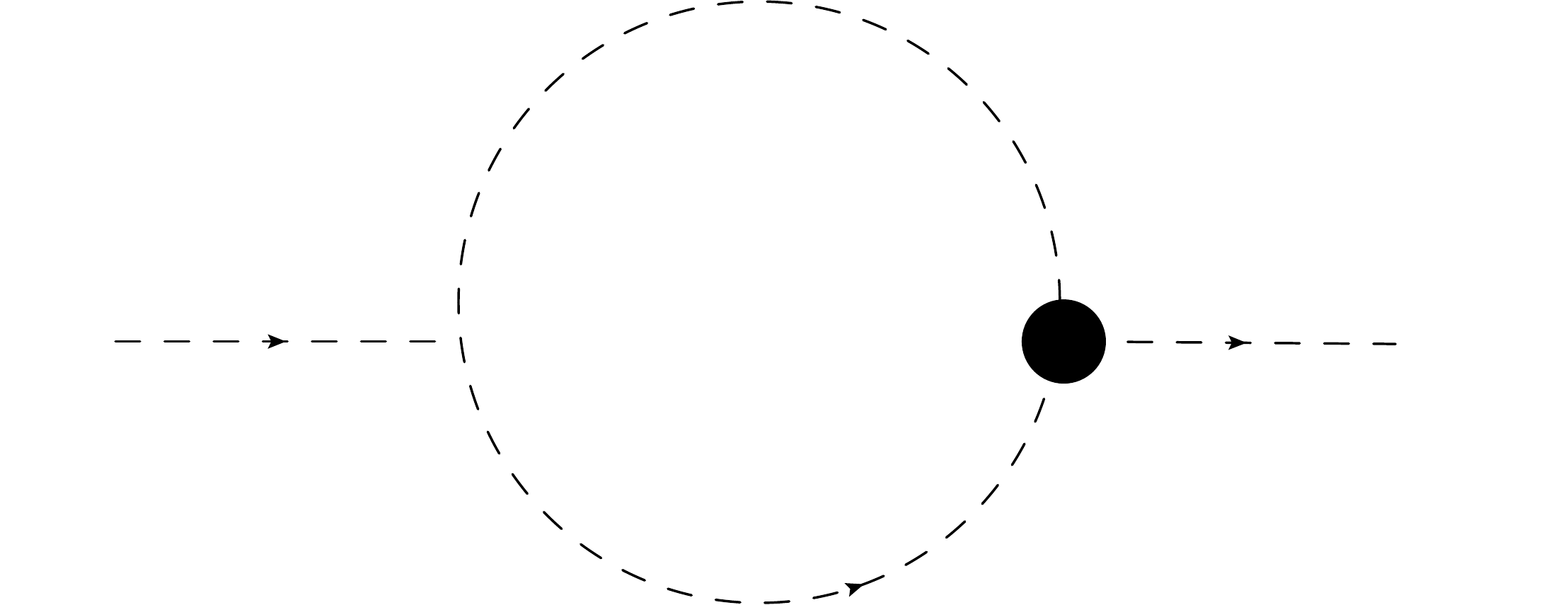}
\includegraphics[width=.45\textwidth]{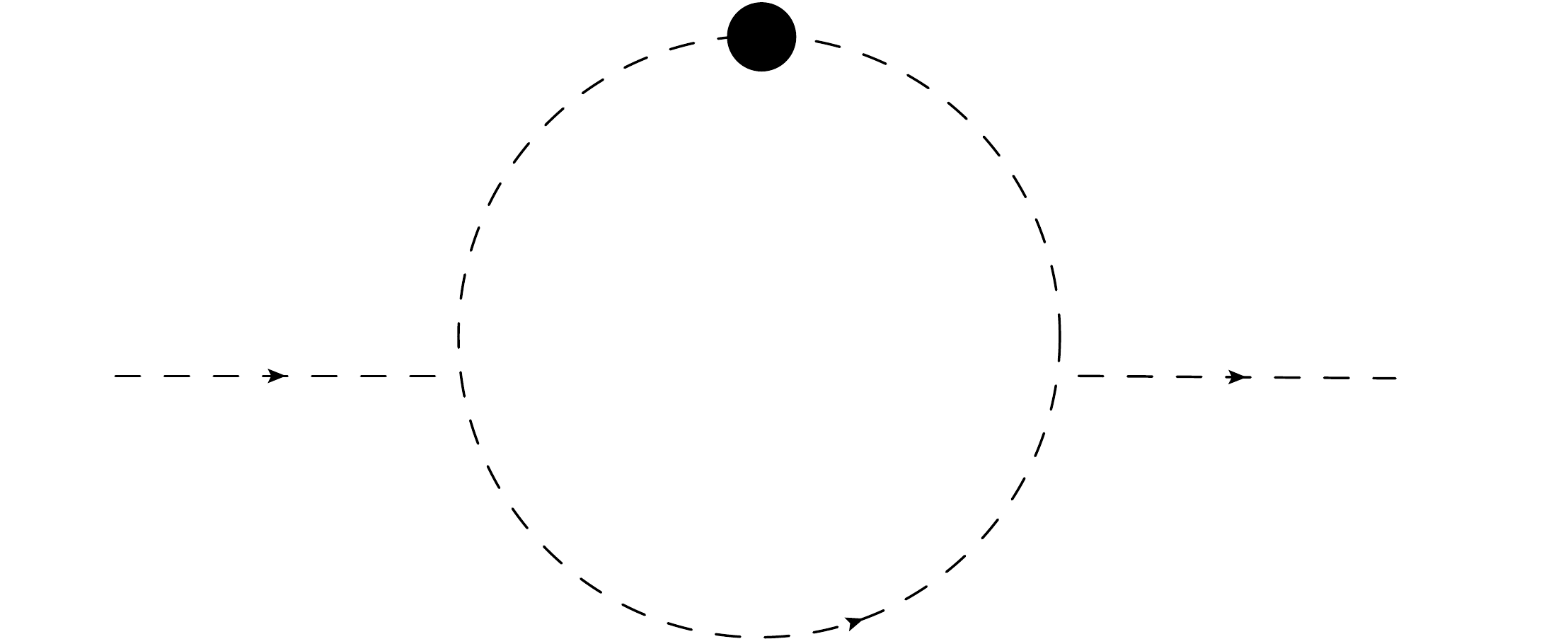}
\caption{\label{2Lmodified} {\small Left: $2$-loop, $2$-point diagram, $\Ga_{2,2a}$, now containing the renormalized 1-loop $3$-point function (dark blob), $\Ga_{3,1}$. Right: Second $2$-loop, $2$-point diagram, $\Ga_{2,2b}$, containing the renormalized $1$-loop $2$-point function (dark blob), $\Ga_{2,1}$.}}
\end{figure}
%%%%%%%%%%%%%%%%%%%%%%%%%%%%%%%%%%%%%%%
\begin{align}
\Ga_{2,3a} & = \frac{i ^ {2}}{2 i ^ {5} M _ {p} ^ {4}} \int \frac{\mathrm{d} ^ 4 k _ {1}}{(2 \pi) ^ 4} \frac{\mathrm{d} ^ 4 k _ {2}}{(2 \pi) ^ 4} \, \frac{V (k _ {1}) V (k_ {2}) V^ {2} (k _ {1}, k _ {2},k _ {3})}{k _ {1} ^ {4}k _ {2} ^ {4}k _ {3} ^ {2}e ^ {2 \kb _ {1} ^ {2}+2 \kb _ {2} ^ {2} + \kb _ {3} ^ {2}}  } \non
&\ra \int \frac{\mathrm{d} ^ 4 k _ {1}}{(2 \pi) ^ 4} \, \frac{V (k _ {1}) \Ga_{3,1r} (k_ {1}, -k _ {1},0) }{k _ {1} ^ {4}e ^ {2 \kb _ {1} ^ {2}}}\,,
\end{align}
where $k_3=-k_1-k_2$.
The key question then is whether this latter integral is finite? A very similar reasoning can be applied to the $2$-loop diagram in Fig.~\ref{fig:2-loop} (right), where one can think of replacing the $2$-point $1$-loop subdiagram with the $1$-loop renormalized $2$-point function, see Fig.~\ref{2Lmodified} (right), and then perform the remaining loop integral:
\begin{align}
\Ga_{2,3b} &= \frac{i ^ {2}}{2 i ^ {5} M _ {p} ^ {4}} \int \frac{\mathrm{d} ^ 4 k _ {1}}{(2 \pi) ^ 4} \frac{\mathrm{d} ^ 4 k _ {2}}{(2 \pi) ^ 4} \, \frac{V^ {2} (k _ {1})  V^ {2}(k _ {1}, -\frac{k _ {1}}{2} + k _ {2},-\frac{k _ {1}}{2}- k _ {2})}{k _ {1} ^ {6}(\frac{k _ {1}}{2} + k _ {2}) ^ {2}(\frac{k _ {1}}{2} - k _ {2}) ^ {2} e ^ { 3 \kb  _ {1} ^ {2}+\LF\frac{\kb _ {1}}{2} + \kb _ {2}\RF ^ {2}+\LF\frac{\kb _ {1}}{2} - \kb _ {2}\RF ^ {2}}}\\
&\ra  \int \frac{\mathrm{d} ^ 4 k_{1}}{(2 \pi) ^ 4} \ \,
\frac{V^ {2} (k_{1})  \Ga_{2,1r}(k_{1},-k_{1})}{k_{1}^ {6}e ^ { 3 \kb _{1}^ {2}}}\,.
\label{eq:43}
\end{align}

\begin{figure}[t]
\centering
\includegraphics[width=.40\textwidth]{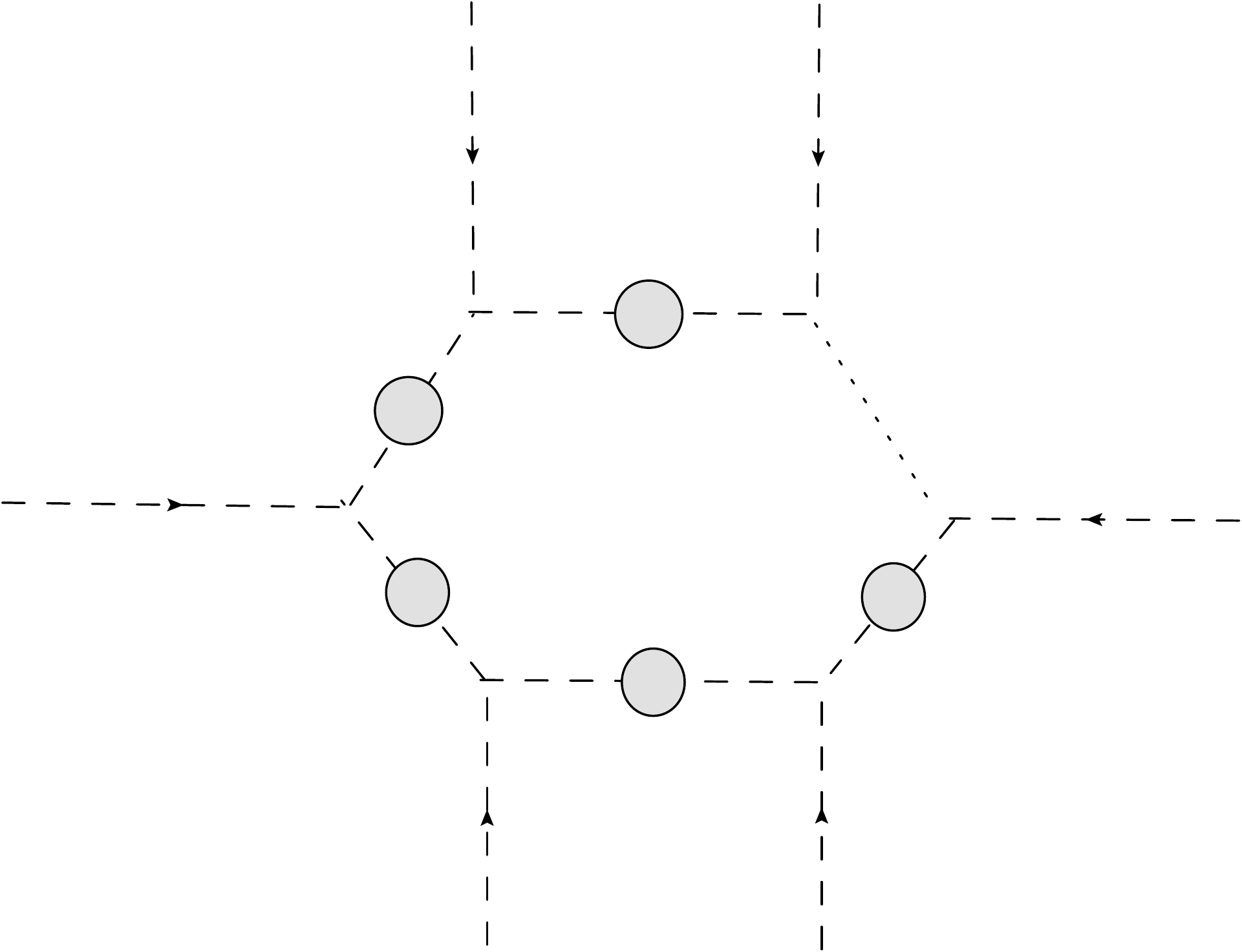}
\hspace{1cm}
\includegraphics[width=.40\textwidth]{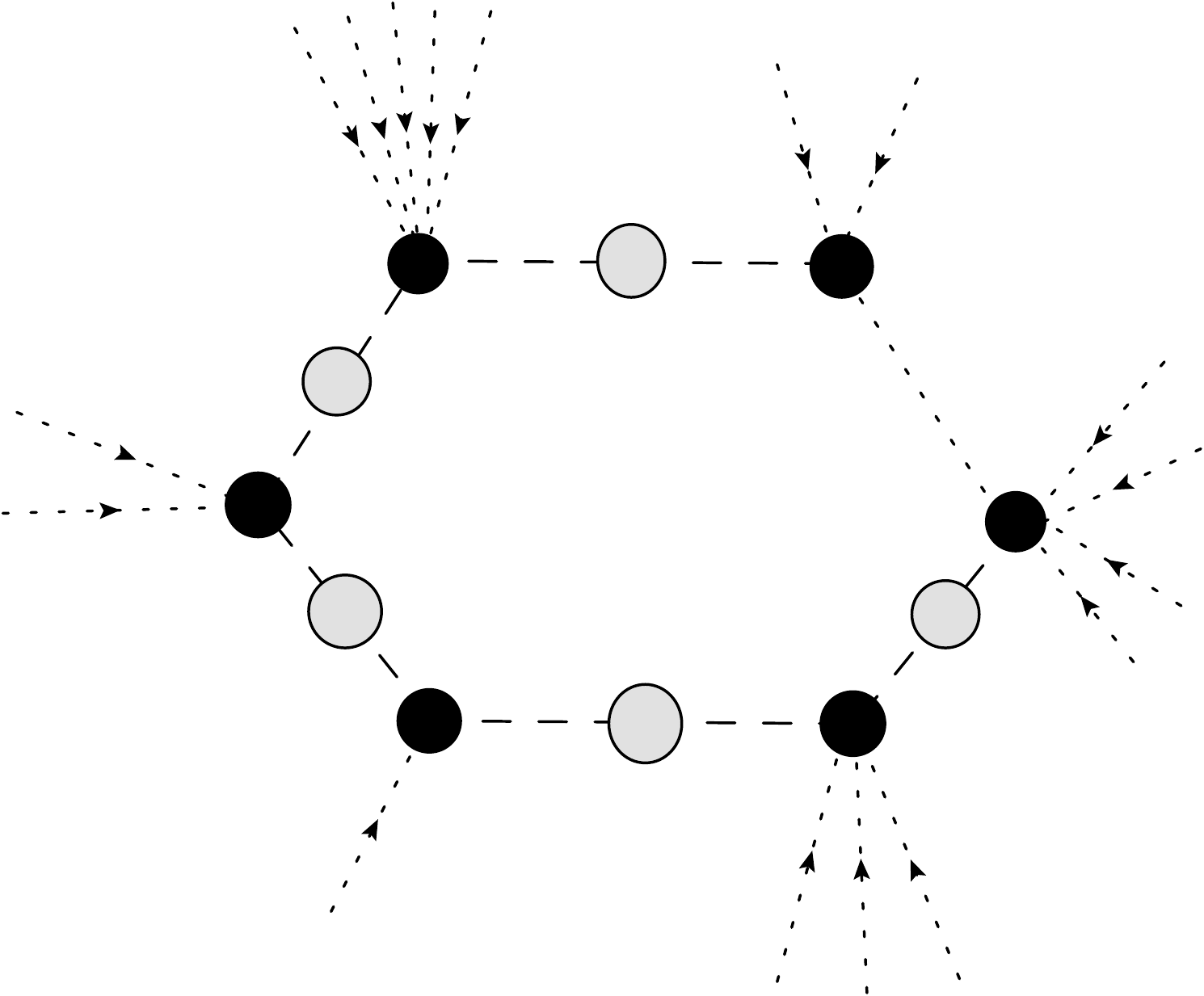}
\caption{\label{fig:Ndressed} {\small Left: $N$-point diagram with dressed propagators (shaded blobs). The dots indicate an arbitrary number of (bare) vertices and dressed propagators. Right: $n$-loop, $N$-point diagram with loop corrections to the vertices (dark blobs) and dressed propagators (shaded blobs). If each dark blob represents an $n_{i}$-loop diagram with $N_i$ external lines, then $n=\sum _ {i} n_{i}+1$ and $N = \sum _ {i} N _ {i}$. The internal dots indicate an arbitrary number of renormalized vertex corrections and dressed propagators. The external dots indicate an arbitrary number of external lines. }}
\end{figure}

Actually, this is a very general prescription, any $n$-loop diagram can be thought of as a $1$-loop integral over a graph containing renormalized vertex corrections and dressed propagators, see Fig.~\ref{fig:Ndressed} (right) for illustration. To prove renormalizability recursively the challenge then is to prove that  if all loops up to $n-1$ are finite, then the remaining $1$-loop integral remains finite too!

Now, we have already seen from counting arguments in in the previous section that if the vertices and the propagators are enhanced and suppressed respectively by the same exponential factor then a $1$-loop diagram remains divergent. This argument can clearly be applied to $n$-loop graphs when viewed as $1$-loop diagrams involving renormalized vertices, and (most importantly) renormalized propagators. What the argument suggests is that to have a chance at renormalizability, the renormalized vertices must  be growing less strongly than the renormalized or the ``dressed'' propagators. In this section, we are  going to compute external momentum dependence of the $2$-point function at $1$-loop, and discuss its ramifications.

%%%%%%%%%%%%%%%%%%%%%%%%%%%%%%%%%%%%%%

\subsection{$1$-loop $2$-point function with arbitrary external momenta}

Calculating the dressed propagator boils down to calculating the $2$-point function with external momenta. At the $1$-loop level
with external momenta $p$,~$-p$ (we assume the convention that the external momenta are incoming and sum to zero), and symmetrical routing of momenta, the Feynman integral is given by (see Appendix~\ref{sec:D1} for details),
\be
\Ga_{2,1}(p^2) = \frac{i}{2 i ^ {2} M _ {p} ^ {2}} \int \frac{d ^ {4} k}{(2 \pi) ^ {4}} \, \frac{V ^ {2} (-p, \frac{p}{2} + k, \frac{p}{2} - k)}{(\frac{p}{2} + k) ^ {2} (\frac{p}{2} - k) ^ {2} e ^ {\LF\frac{\pb}{2} + \kb\RF ^ {2}} e ^ {\LF\frac{\pb}{2} - \kb\RF ^ {2}}} \,,
\label{2pt-ext}
\ee
where
\begin{align}
V^2 \LF -p, \frac{p}{2} + k, \frac{p}{2} - k \RF & = i ^ {2} C ^ {2}\LT 1 - 2  e ^ {\LF\frac{\pb}{2} + \kb\RF ^ {2}} - 2  e ^ {\LF\frac{\pb}{2} - \kb\RF ^ {2}} - 2  e ^ {\pb ^ {2}} + 2  e ^ {\LF\frac{\pb}{2} + \kb\RF ^ {2}} e ^ {\LF\frac{\pb}{2} - \kb\RF ^ {2}} \Rd \non
& + \Ld 2  e ^ {\LF\frac{\pb}{2} - \kb\RF ^ {2}} e ^ {\pb ^ {2}} + 2  e ^ {\pb ^ {2}} e ^ {\LF\frac{\pb}{2} + \kb\RF ^ {2}} +  e ^ {2\LF\frac{\pb}{2} + \kb\RF ^ {2}} +  e ^ {2\LF\frac{\pb}{2} - \kb\RF ^ {2}}
+  e ^ {2 \pb ^ {2}} \RT \\ \label{Vextsq}
& = i ^ {2} C ^ {2}\LT 1 - 2  e ^ {\LF\frac{\pb}{2} + \kb\RF ^ {2}} - 2  e ^ {\LF\frac{\pb}{2} - \kb\RF ^ {2}} - 2  e ^ {\pb ^ {2}} + 4  e ^ {\LF\frac{\pb}{2} + \kb\RF ^ {2}} e ^ {\LF\frac{\pb}{2} - \kb\RF ^ {2}} \Rd \non
& +  2  e ^ {\LF\frac{\pb}{2} - \kb\RF ^ {2}} e ^ {\pb ^ {2}} + 2  e ^ {\pb ^ {2}} e ^ {\LF\frac{\pb}{2} + \kb\RF ^ {2}} + \LF e ^ {2\LF\frac{\pb}{2} + \kb\RF ^ {2}}-e ^ {\LF\frac{\pb}{2} + \kb\RF ^ {2}} e ^ {\LF\frac{\pb}{2} - \kb\RF ^ {2}}\RF \non
& + \Ld  \LF e ^ {2\LF\frac{\pb}{2} - \kb\RF ^ {2}}-e ^ {\LF\frac{\pb}{2} + \kb\RF ^ {2}} e ^ {\LF\frac{\pb}{2} - \kb\RF ^ {2}}\RF
+  e ^ {2 \pb ^ {2}} \RT
\end{align}
and
\be
C = \frac{1}{4} \LT p ^ {2} + \LF \frac{p}{2} + k\RF ^ {2} + \LF \frac{p}{2} - k\RF ^ {2}\RT\,.
\ee
While using a cut-off scheme to regulate the integral is more instructive to see the divergent structure, technically it is much more convenient to use dimensional regularization, which is what we will employ from here onwards. The integral in Eq.~\eqref{2pt-ext} contains several terms coming from the various sums of exponents that make up the vertex functions. The different integrals arising from the sum in $V^2$ can be grouped in three ways:
\begin{enumerate}
\item[(I)] When the integrand contains no exponentials, this comes from the fifth term in Eq.~(\ref{Vextsq}), and gives a divergent result.
\item[(II)] When we have a Gaussian damping term present in Eq.~(\ref{Vextsq}). This is the case for all the terms except the fifth, eighth and the ninth terms and gives a convergent answer.
\item[(III)] The eighth and the ninth terms in Eq.~\eqref{Vextsq} give rise to integrals containing the terms $e^{2 \pb \cdot \kb}-1$ and $e^{-2 \pb \cdot \kb}-1$, respectively, but they are not particularly  important for our discussion as  will become clear soon. Let us discuss these terms separately now.
\end{enumerate}
%%%%%%%%%%%%%%%%%%%%%%
\subsubsection{Group (I) terms}
The divergent integral corresponding to the fifth term can again be calculated straightforwardly using dimensional regularization (see Appendix~\ref{sec:DR}), and one obtains
\be
\Ga_{2,1,i}(p^2)=\frac{i p ^ {4}}{128 \pi ^ {2} M _ {p} ^ {2}} \left(\frac{2}{\epsilon} - \log \left(\frac{p^2}{4 \pi M^2}\right) - \gamma +2 \right)\,.
\ee
Let us make a couple of comments: Firstly, the $p^{2}\ra 0$ limit is well defined, {\it i.e.}, none of the expressions diverge. If it did, that would make the low energy limit ill-defined, ruling out such modifications phenomenologically  even as an effective theory. Secondly, the counter term needed to cancel the divergence is given by
\be
\cL_{\mt{ct}}= - \frac{1}{128 \en \pi ^{2}M_{p}^{2}}\int d^{4}x \, \phi \Box^2 \phi \,,
\ee
or, equivalently,
\be
\Ga_{2,1,\mt{ct}}(p^2)=-\frac{i p ^ {4}}{64 \pi ^ {2} M _ {p} ^ {2}} \frac{1}{\epsilon}\,.
\ee
The counter term is clearly not of the same form as the original action Eq.~\eqref{eq:action}. This is not surprising given that the symmetry principle we used to write down the action Eq.~\eqref{eq:action} was not a symmetry of the action, but only that of the field equations.

%%%%%%%%%%%%%%%%%%%%%%%%%%%%%%%%%%%%%%%%%%%%%%%%%
\subsubsection{Group (II) terms}
The group (II) type integrals are all convergent due to the presence of the exponential damping factor. They can therefore be evaluated rather straightforwardly to yield
\begin{align}
&\Ga_{2,1,ii}(p^2)  = \frac{i M^4 e^{-\pb^2}}{512 M _ {p}^{2}\pi ^2 \pb^2}\LT \vphantom{\text{Ei}\LF \frac{\pb^2}{2}\RF}-2 e^{\pb ^ 2} \LF e ^ {2 \pb ^2} - 1 \RF \pb^6 Ei\LF-\pb^2 \RF + \LF e ^ {\pb^2} - 1\RF \LF \vphantom{\text{Ei}\LF \frac{\pb^2}{2}\RF}-2 \LF \pb^4 + 3 \pb^2 + 2 \RF \Rd \Rd \non
& + \Ld \Ld \LF e^{\frac{3 \pb^2}{2}} -e^{\frac{\pb^2}{2}} \RF \LF 2\pb^4 + 5 \pb^2 + 4 \RF + e ^{\pb^2} \LF e^{\pb^2} -1 \RF \pb^6 Ei \LF - \frac{\pb^2}{2} \RF + 2 e^{\pb^2} \LF 7 \LF \pb^4 + \pb^2 \RF +2 \RF \RF \RT.
\end{align}
Again, the expression is regular as $p^{2} \ra 0$. This again shows that the theory has a well defined low energy limit. However, to assess the renormalizability of the theory we need to look at the UV behavior, and especially track any exponential growth. With this in mind let us look at the various terms that grow as a Gaussian as $p^2 \ra \infty$:
\begin{align}
\Ga_{2,1,ii}(p^2) & =\frac{i M^2 }{512 M _ {p} ^ {2} \pi ^2 p^2} \left[ e^{\frac{3\bar{p}^2}{2 }} \left(4 M^4+5 M^2 p^2+2 p^4\right)+2 e^{\bar{p}^2} \left(2 M^4+7 M^2 p^2+7 p^4\right) \Rd \non
& - \Ld 2 e^{\frac{\bar{p}^2}{2}}\left(4 M^4+5 M^2 p^2+2 p^4\right) \vphantom{e^{\frac{p^2}{M^2}}} \right]
- \frac{iM^2p ^ {2} }{256 M _ {p} ^ {2} \pi ^2 } \left[\left(1-2M^{2}p^{-2}+8M^{4}p^{-4}\right)e^{\frac{3\bar{p}^2}{2}}-2e^{\frac{\bar{p}^2}{2}} - e^{\bar{p}^2}  \right]\non
& +\dots \,,
\end{align}
where the $\dots$ indicate subleading terms or terms which are growing at most as a polynomial, and we have used  the relation,
\be
\lim _ {x \to + \infty} x ^ {2} e ^ {\alpha x ^ {2}} Ei \left(- \alpha x ^ {2} \right) = - \frac{1}{\alpha}\,,
\ee
for positive $\alpha$ to obtain the asymptotic behavior. In particular, as $p^2\ra \infty$, we find
\be
\Ga_{2,1,ii}(p^2)\ra\frac{i M^4 e^{\frac{3\pb^2}{2}}}{512M _ {p} ^ {2} \pi ^2}\LT  9-12\pb^{-2}+\dots \RT\,.
\ee
As we see, the correction to the propagator grows with a larger exponent than the ``bare'' inverse propagator, and this will be crucial in proving finiteness of the $1$-loop diagrams and our arguments on renormalizability of the theory.

%%%%%%%%%%%%%%%%%%%%%%%%%%%%%%%%%%%%%%%%%%%%%%%%%
\subsubsection{Group (III) terms}

%%%%%%%%%%%%%%%%%%%%%%%%%%%%%%%%

For the purpose of completeness, let us also compute the group (3) integrals using dimensional regularization (see Appendix~\ref{sec:DR} for details). We find that the $e ^ {2\pb \cdot \kb}$ integrals do not give rise to any poles. In fact, we have that (see Appendix~\ref{sec:DR2})
\be
\label{eq:p.k}
\Ga_{2,1,iii}(p^2) = 0 \,.
\ee
As we can see, the $\Ga_{2,1,ii}(p^2)$ term dominates for large momentum, and is therefore going to be the most important for understanding the UV behavior of the quantum theory.

To summarize, from our preceding calculations   in the UV limit we have:
\be
\Ga_{2,1}(p^2)=\Ga_{2,1,i}(p^2) + \Ga_{2,1,ii}(p^2)+\Ga_{2,1,iii}(p^2) \approx \frac{9 i M^4 e^{{3\bar{p}^2}/{2}}}{512 M _ {p} ^ {2} \pi ^2}\,.
\ee
In other words, the $2$-point ``vertex'' grows more strongly than even the  momentum dependence, $\sim e^{\bar{p}^2}$,  of the bare $3$-point vertex. Also note that the term is finite and therefore
it  is expected to survive even after we have renormalized the divergent part in the $1$-loop $2$-point function. Naively, this may seem like a disaster. For instance, it is easy to see that this leads to an additional divergence in the $2$-loop diagram, Fig.~\ref{2Lmodified} (right), which contains the $1$-loop $2$-pt subdiagram. In Eq.~\eqref{eq:43}, since $\Ga_{2,1r}(k^2)$ goes as $e^{\frac{3\kb^2}{2}}$, the integrand now diverges exponentially as $e^{\frac{\kb^2}{2}}$. This is worse than the power law divergence of $1$-loop! So, does this mean the end of the road for non-local theories as a candidate for quantum gravity? On the contrary, we will now see that this apparent strong exponential dependence may, in fact, be exactly what is needed to make all the higher loops finite!

%%%%%%%%%%%%%%%%%%%%%%%%%%%%%%%%%%%%%%%%%%%%%%%%%%%%%%%%%%
\section{Improved convergence with dressed propagators}\label{sec:dressed}
\numberwithin{equation}{section}
%%%%%%%%%%%%%%%%%%%%%%%%%%%%%%%%%%%%%%%%%%%%%%%%%%%%%%%%%%%
%%%%%%%%%%%%%%%%%%%%%%%%%%%%%%%%%
\subsection{Dressed propagator \& $1$-loop integrals}
\begin{figure}[t]
\centering
\includegraphics[width=.55\textwidth]{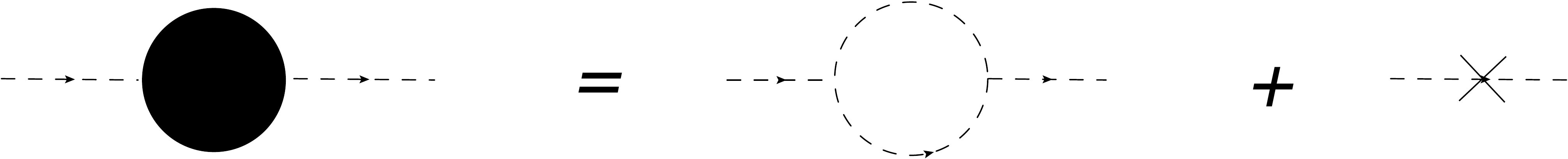}
\vspace{1cm}\\
\includegraphics[width=1.0\textwidth]{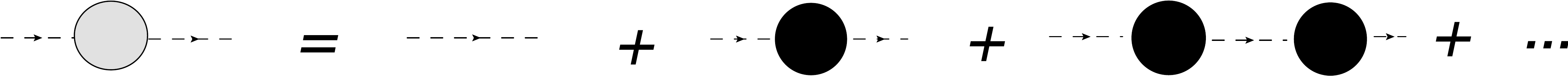}
\caption{\label{fig:dressed} {\small Top: The $1$-loop, $2$-point contribution of $1$PI diagrams. The cross denotes a counter term vertex. Bottom: The dressed propagator as the sum of an infinite geometric series. The dressed propagator is denoted by the shaded blob.}}
\end{figure}

We saw in the earlier section that additional divergences may arise in higher loops from the $1$-loop, $2$-point functions.
However, we know that, in quantum field theory, the $1$-loop correction is only the first term in a sequence of graphs, see Figs.~\ref{fig:dressed}, which can be resummed as a geometric series in the region of convergence and then analytically continued to the entire momentum space. In other words, the bare propagators need to be replaced by the dressed propagator while performing calculations for higher-point Green's functions or higher loops. Note, that no such infinite sequence exists for interaction vertices, the loop contributions simply add to the bare vertex. Thus, a rather remarkable consequence of this resummation will be that for non-local theories the dressed propagators will be more exponentially suppressed than their bare counterparts at large momentum, and therefore going to overwhelm the exponential enhancements coming from the vertices~\footnote{This property is more general than just the infinite-derivative theories as finite 1-loop results were also obtained for ``local'' higher-derivative extensions of gravity~\cite{Donoghue:1999qh}.}. In particular, we will explicitly see that this will make the UV part of all higher (than two) point $1$-loop graphs finite!

The UV part of the $2$-loop integrals, and here we will only illustrate the $2$-point function, will also become finite. We will  argue that it should be possible to extend this reasoning to all higher loop graphs. In other words, we conjecture that except for the $1$-loop, $2$-point function, all graphs in  this toy-model for quantum gravity  converge in the UV. A more rigorous proof, possibly involving more general coupling (not just cubic), is beyond the scope of this paper and obviously requires further investigation.

%%%%%%%%%%%%%%%%%%%%%%%%%%
%%%%%%%%%%%%%%%%%%%%%%

The $1$-loop, $2$-point contribution schematically reads (see Fig.~\ref{fig:dressed}, top,  for a diagrammatic representation):
\be
\Ga_{2,1}(p^2)+\Ga_{2,1,\mt{ct}}(p^2)=\Ga_{2,1\mt{r}}(p^2)= \frac{i M^4}{M_p^2}f(\pb^2)\,,
\ee
where
\begin{align}
f (\pb^2) & = \frac{\pb ^ {4}}{128 \pi ^ {2}} \left(- \log \left(\frac{\pb^2}{4 \pi}\right) - \gamma +2 \right) \non
& + \frac{e^{-\pb^2}}{512 \pi ^2 \pb^2} \LT \vphantom{\text{Ei}\LF \frac{\pb^2}{2}\RF}-2 e^{\pb ^ 2} \LF e ^ {2 \pb ^2} - 1 \RF \pb^6 Ei \LF-\pb^2 \RF + \LF e ^ {\pb^2} - 1\RF \LF \vphantom{\text{Ei}\LF \frac{\pb^2}{2}\RF}-2 \LF \pb^4 + 3 \pb^2 + 2 \RF \Rd \Rd \non
& + \Ld \Ld  \LF e^{\frac{3 \pb^2}{2}} -e^{\frac{\pb^2}{2}} \RF \LF 2\pb^4 + 5 \pb^2 + 4 \RF + e ^{\pb^2} \LF e^{\pb^2} -1 \RF \pb^6 Ei \LF - \frac{\pb^2}{2} \RF + 2 e^{\pb^2} \LF 7 \LF \pb^4 + \pb^2 \RF +2 \RF \RF \RT \,.
\end{align}
$f(\pb^2)$ is a regular analytic function of $\pb^2$ which grows as $e^{3\pb^2/2}$ as $p^2 \ra \infty$. One can observe that $f(\pb^2) \ra 0$ as $p^{2} \ra 0$. Hence, $\Ga_{2,1\mt{r}}(p^2) \ra 0$ as $p^{2} \ra 0$, thereby implying that the renormalized $1$-loop, $2$-point function has a well defined low-energy limit. The dressed propagator then represents the geometric series of all the graphs with $1$-loop, $2$-point insertions as shown in Fig.~\ref{fig:dressed} (bottom), analytically continued to the entire complex $p^2$-plane.  Mathematically, this is equivalent to replacing the bare propagator, $\Pi(p^2)$, with the dressed propagator, $\w{\Pi}(p^2)$:
\be
\w{\Pi}(p^2)= \frac{\Pi(p^2)}{1-\Pi(p^2)\Ga_{2,1\mt{r}}(p^2)}= \frac{-i}{p ^ {2 } e ^ { \bar{p} ^ {2}}-\frac{M^4}{M_p^2}f\LF\bar{p}^2\RF}\,.
\ee
Since, in our case, $\Pi(p^2)\Ga_{2,1\mt{r}}(p^2)$ grows with large momenta, in the UV limit, we have
\be
\label{dressed-UV}
\w{\Pi}(p^2)\ra \Ga^{-1}_{2,1\mt{r}}(p^2)\approx \LF 9-12 \pb^{-2}  \RF ^ {-1} e^{-\frac{3\bar{p}^2}{2}}\,.
\ee
Clearly, the dressed propagator is more strongly suppressed than the bare propagator! This is a very crucial result that is now going to  ensure that the UV contribution of quantum fluctuations for all the other higher-point $1$-loop graphs are finite. Since the cubic interactions are known to cause vacuum instability,  the mass squared  correction is negative (as is the case in ordinary $\phi^3$ theory) leading to an artificial pole in the dressed propagator. This makes all the integrals divergent  in the IR. We however, expect  this pathology to be cured once higher-order (such as quartic couplings) interactions are included as they must be in a complete gravitational theory. Here, we are concentrating on the UV behavior, and we can bypass this problem by  evaluating only the UV part of the integrals, from say $\pb= 1\dots \infty$ where the dressed propagator can be approximated by Eq. (\ref{dressed-UV}).

\begin{figure}[t]
\centering
\includegraphics[width=.40\textwidth]{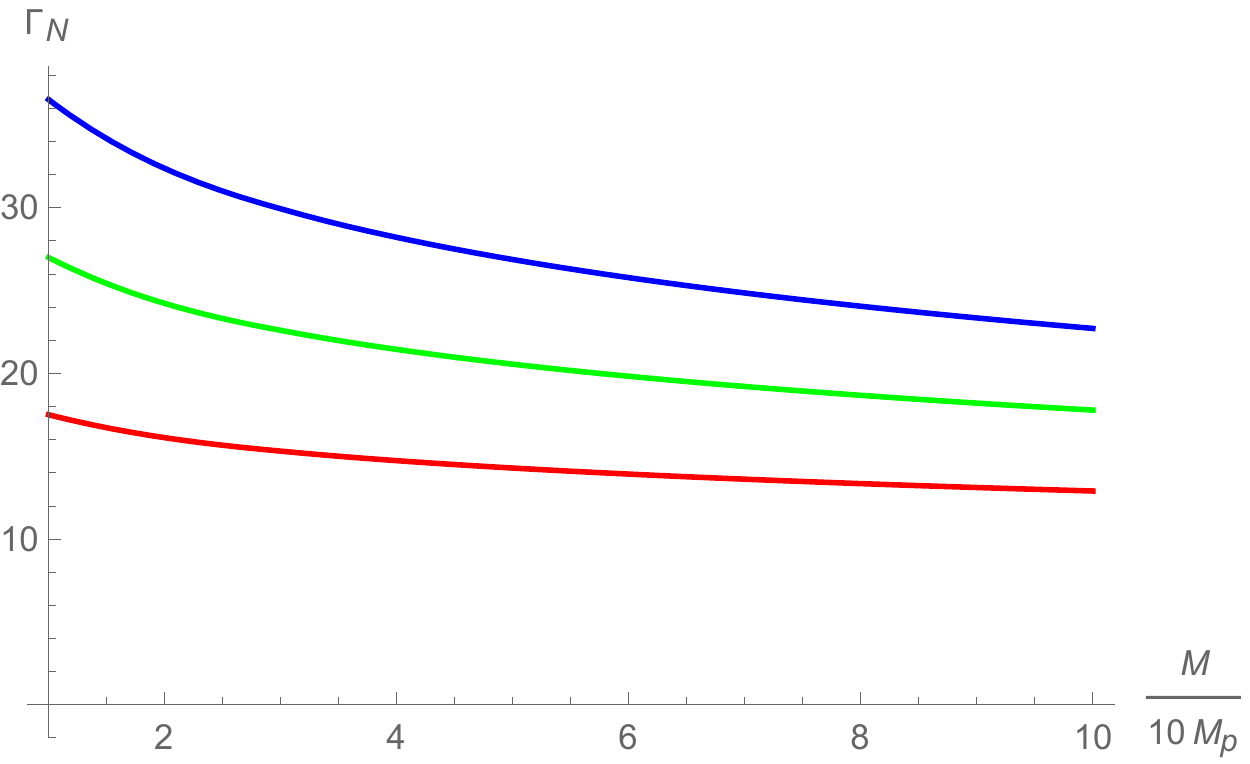}
\caption{\label{fig:nine} {\small A log-plot for $3$-point diagrams $\Ga _ {3}$ (in units of $i M_{p}$), $4$-point diagrams $\Ga _ {4}$ (in units of $i$) and $5$-point diagrams $\Ga_{5}$ (in units of $i M_{p}^{-1}$) where ${M}/{M _ {p}}$ ranges from $0.1$ to $1$. The red, green and blue curves represents Eq.~\eqref{eq:N,UV} for $N = 3,4$ and $5$, respectively.}}
\end{figure}

Let us now revisit the $1$-loop calculations of the $N$-point diagrams. Once the infinite sum of diagrams leading to the dressed propagators are taken into account,
see Fig.~\ref{fig:Ndressed} (left), the  UV part of the $1$-loop integral reduces to
\be
\label{eq:N,UV}
\Ga_{N,UV} \approx \frac{i}{i ^ {N} M _ {p} ^ {N}}\int \frac{\mathrm{d} ^  4 k}{(2 \pi) ^ 4} \, \frac{V ^ {N}(k)}{\LT -\frac{M^4}{M_p^2}f\LF\kb\RF\RT^N}=\frac{i M ^ {4}}{M _ {p} ^ {N}}\int \frac{\mathrm{d} ^ 4 \kb}{(2 \pi) ^ 4} \, \frac{\kb ^ {2 N} e ^ {N \kb^2}}{\LT \frac{M^2}{M_p^2}f\LF{\kb}\RF\RT^N}\,.
\ee
This integral is finite and we have provided numerical plots as a function of $M/M_p$, see Fig.~\ref{fig:nine}. We note that the amplitudes remain well behaved even in the limits $M\ll M_p$ and $M_p\ll M$.
%%%%%%%%%%%%%%%%%%%%%%%%%%%%%%%%%%%%%%%
\subsection{UV convergence of $2$-loop diagrams}\label{sec:with-momentum}
%%%%%%%%%%%%%%%%%%%%%%%%%%%%%%%%%%%%%%%%%%%%
In the previous section we have seen how when we make the transition from the bare to the dressed propagator, the $1$-loop diagrams in non-local theories become finite. This means that to renormalize  at the $1$-loop level, all we have to do is to renormalize the divergence in the $2$-point function. This is not very different from the local field theories. For instance, in $\la\phi^4$ theory, once the $2$-point and $4$-point functions are renormalized, all the higher-point Green's functions become finite. Can this procedure be extended to all loops though? If the answer is yes, it would provide a tremendous encouragement towards the possibility of having a candidate for an analogue of quantum theory of gravity in the non-local framework. However, in this paper we have to be content with mostly a study of the convergence properties of  the $2$-point $2$-loop diagrams. We will indeed see that they also become finite as opposed to the $\La^4$ divergence found in Eq.~\eqref{eq:6666}.

\begin{figure}[t]
\centering
\includegraphics[width=.40\textwidth]{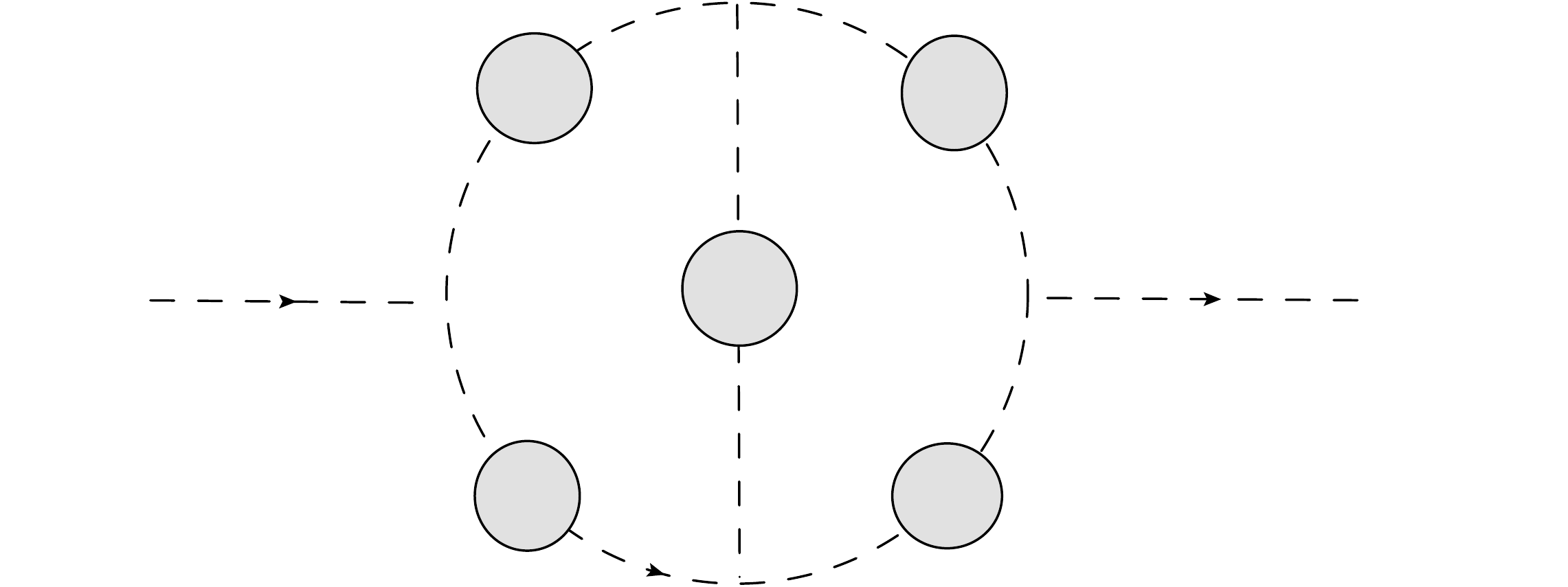}
\includegraphics[width=.40\textwidth]{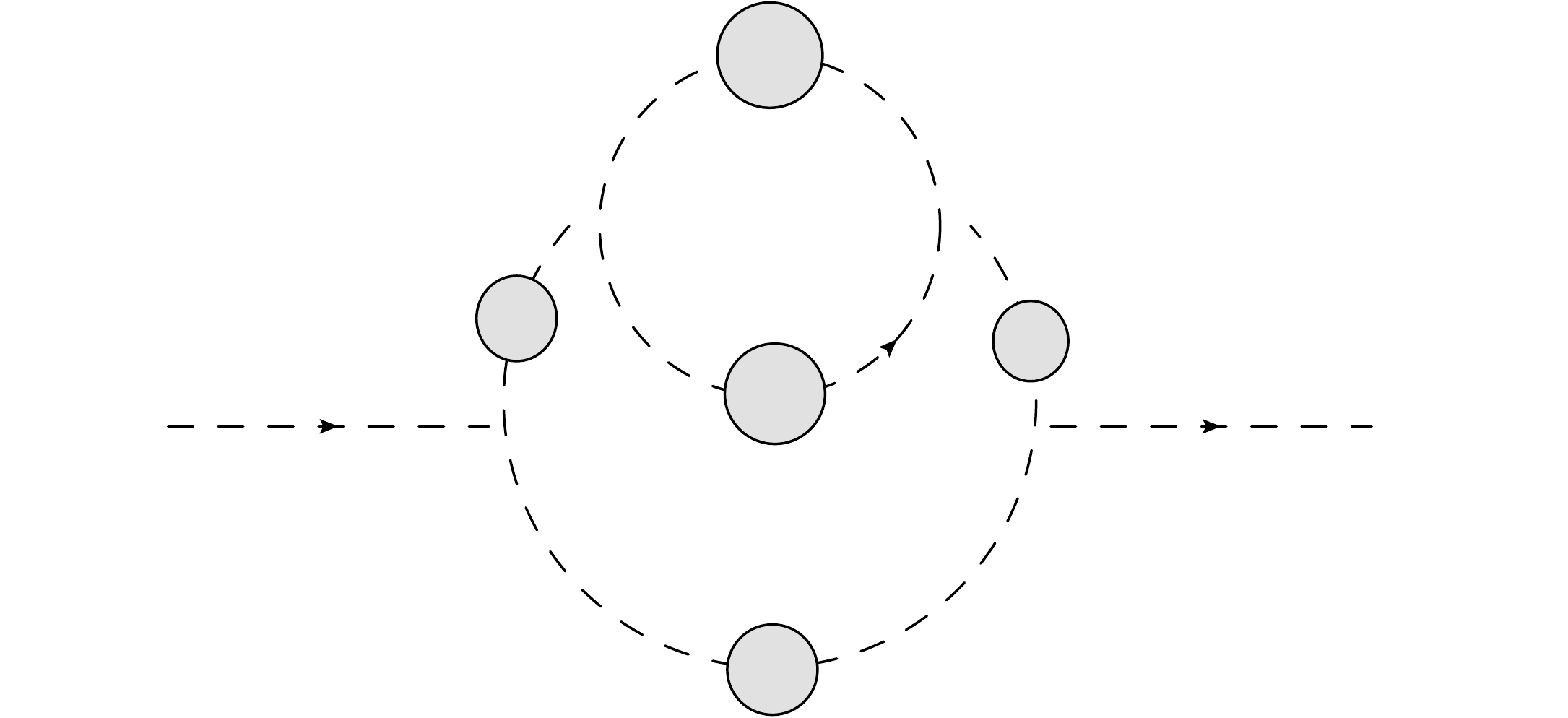}
\caption{\label{fig:2Ldressed} Left: The $2$-loop, $2$-point diagram $\Gamma_{2,3a}$. The shaded blobs denote dressed propagators. Right: The $2$-loop, $2$-point diagram $\Ga_{2,3b}$. The shaded blobs denote dressed propagators.}
\end{figure}

Consider first the Fig.~\ref{fig:2Ldressed} (right) that resembles the $1$-loop, $2$-point Fig.~\ref{fig:2-loop} (right), except  that now the bare propagator has been replaced by the dressed propagator. Again, to determine the finiteness of the graph it is sufficient to focus on the zero external momenta case. The Feynman integral is given by
%%%%%%%%%%%%%%%%%%%%%%%%%%%%%
%%%%%%%%%%%%%%%%%%%%%%%%%%%%%%%%%
\begin{align}
\label{eq:58}
\Ga_{2,3b} & = \frac{i ^ {2}}{2 i ^ {5} M _ {p} ^ {4}} \int \frac{\mathrm{d} ^ 4 k _ {1}}{(2 \pi) ^ 4} \frac{\mathrm{d} ^ 4 k _ {2}}{(2 \pi) ^ 4} \, \frac{V^ {2} (k _ {1})  V^ {2}(k _ {1}, -\frac{k _ {1}}{2} + k _ {2},-\frac{k _ {1}}{2}- k _ {2})}{\LT k _ {1} ^ {2}e ^ {\kb_{1}^{2}}-\frac{M^4}{M_p^2}f\LF\kb _{1}^{2}\RF\RT^3 \LT(\frac{k _ {1}}{2} + k _ {2}) ^ {2}e ^ {\LF\frac{\kb _ {1}}{2} + \kb _ {2}\RF ^ {2}}-\frac{M^4}{M_p^2}f\LF\LF\frac{\kb _ {1}}{2} + \kb _ {2}\RF ^ {2}\RF\RT}\non
& \times \frac{1}{\LT (\frac{k _ {1}}{2} - k _ {2}) ^ {2}   e ^ {\LF\frac{\kb _ {1}}{2} - \kb _ {2}\RF ^ {2}}-\frac{M^4}{M_p^2}f\LF\LF\frac{\kb _ {1}}{2} - \kb _ {2}\RF ^ {2}\RF\RT}\,. \quad
\end{align}
Making the redefinition $k _ {1} \to k _ {1}$, $- \frac{k _ {1}}{2} - k _ {2}\to k _ {2}$ and $k_3=-k_1-k_2$, we get
\be
\label{eq:59}
\Ga_{2,3b} = \frac{1}{2 i^{3}  M _ {p} ^ {4}} \int \frac{\mathrm{d} ^ 4 k _ {1}}{(2 \pi) ^ 4} \frac{\mathrm{d} ^ 4 k _ {2}}{(2 \pi) ^ 4}\frac{V^ {2} (k _ {1})  V^ {2}(k _ {1}, k _ {2},k _ {3})}{\LT k _ {1} ^ {2}e ^ {\kb_{1}^{2}}-\frac{M^4}{M_p^2}f\LF\kb _{1}^{2}\RF\RT^3 \LT k _ {2} ^ {2}e ^ {\kb _ {2} ^ {2}}-\frac{M^4}{M_p^2}f\LF\kb _ {2} ^ {2}\RF\RT\LT \kb _ {3} ^ {2}   e ^ {\kb _ {3} ^ {2}}-\frac{M^4}{M_p^2}f\LF\kb _ {3} ^ {2}\RF\RT}\,.
\ee
To see whether the integrals are convergent or not, we need to look at large values of $k_1,~k_2$ where the exponentials dominate. One can check that the integrand goes as
%\be
%\sim \exp\LT -(k_1+k_2)^2-{7\over 4}k_1^2\RT\ ,
%\ee
\be
\sim \exp\LT -(\frac{k_1}{2}-k_2)^2-\frac{7}{4}k_1^2\RT\,,
\ee
ensuring that both the $k_1$ and $k_2$ integrals are convergent.
%One can compute these integrals numerically, and we have provided plots %as a function of the parameter $M/M_p$.

The other $2$-loop diagram for the $2$-point function, see Fig.~\ref{fig:2Ldressed} (left), reads as:
\be
\Ga_{2,3a} = \frac{i ^ {2}}{2 i ^ {5} M _ {p} ^ {4}} \int \frac{\mathrm{d} ^ 4 k _ {1}}{(2 \pi) ^ 4} \frac{\mathrm{d} ^ 4 k _ {2}}{(2 \pi) ^ 4}\,   \frac{V (k _ {1}) V (k_ {2}) V^ {2} (k _ {1}, k _ {2},k _ {3})}{\LT k _ {1} ^ {2}e ^ {\kb_{1}^{2}}-\frac{M^4}{M_p^2}f\LF\kb_{1}^{2}\RF\RT^2\LT k _ {2} ^ {2}e ^ {\kb_{2}^{2}}-\frac{M^4}{M_p^2}f\LF\kb_{2}^{2}\RF\RT^2 \LT k _ {3} ^ {2}e ^ {\kb_{3}^{2}}-\frac{M^4}{M_p^2}f\LF\kb_{3}^{2}\RF\RT }\,,
\ee
where $k_3=-k_1-k_2$. The exponential dependence of the integrand as $|k_1|,|k_2|\ra \infty$, goes as
\be
\sim \exp\LT -\frac{3}{2}\LF k_1-\frac{k_2}{3}\RF^2-\frac{4}{3}k_2^2\RT\,,
\ee
again leading to a convergent integral.

%%%%%%%%%%%%%%%%%%%%%%%%%%%%%%

\subsection{Higher vertices and prospects for a finite theory}
\begin{figure}[t]
\centering
\includegraphics[width=.40\textwidth]{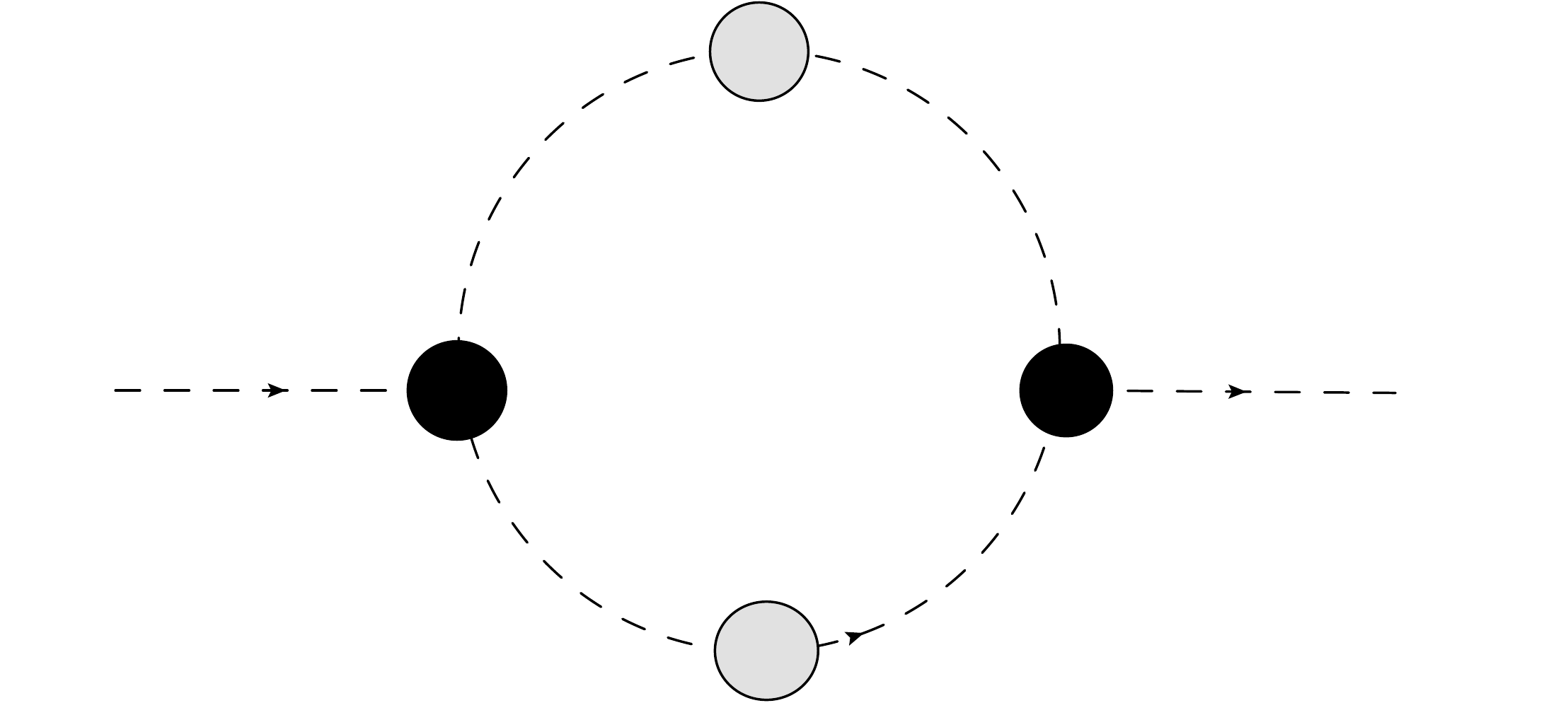}
\includegraphics[width=.40\textwidth]{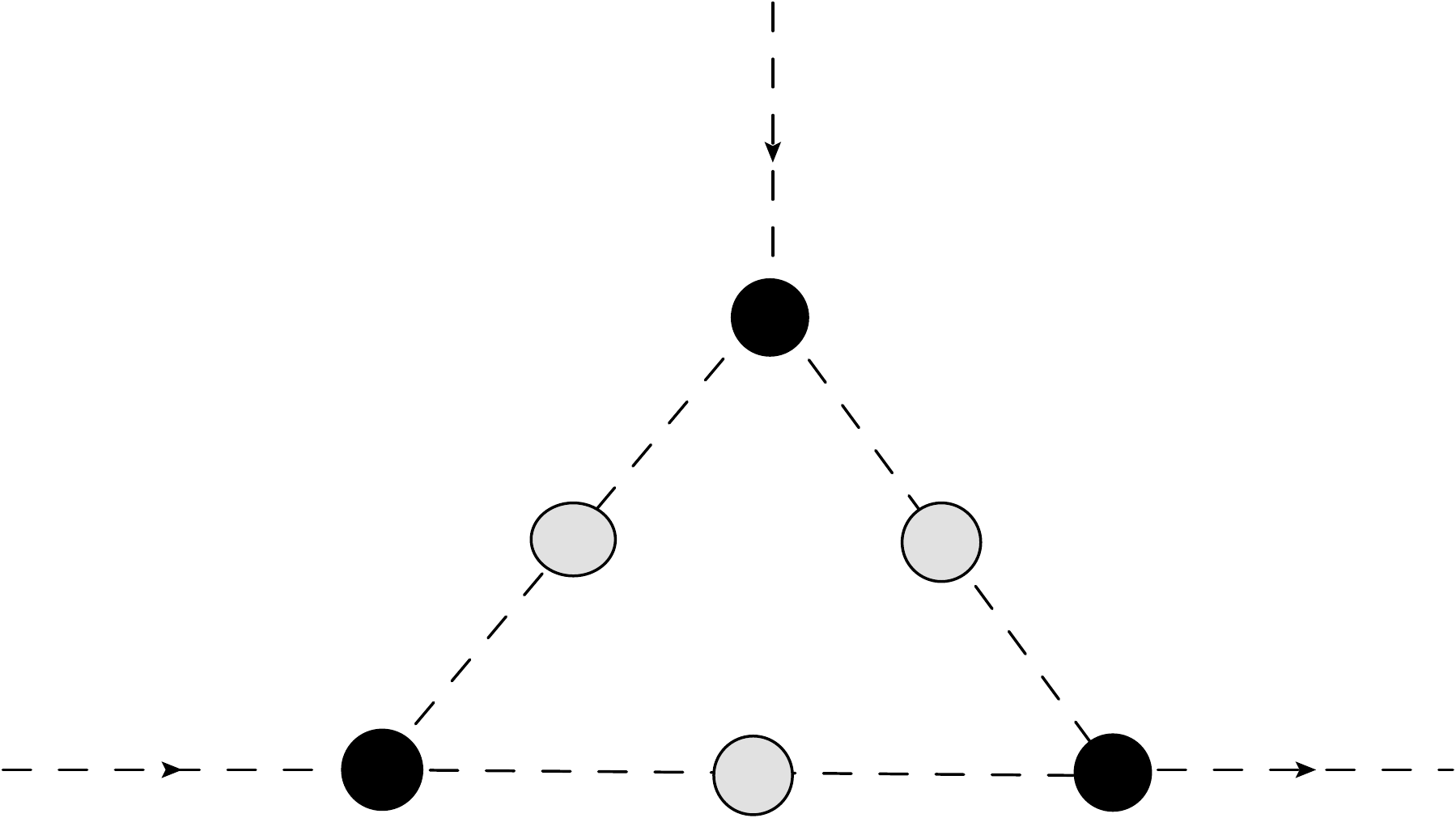}
\caption{\label{set} {\small Left: $2$-point diagram constructed out of lower-loop $2$-point \& $3$-point diagrams. The shaded blobs indicate dressed propagators and the dark blobs indicate renormalized vertex corrections. Right: $3$-point diagram constructed out of lower-loop $2$-point \& $3$-point diagrams. The shaded blobs indicate dressed propagators and the dark blobs indicate renormalized vertex corrections.}}
\end{figure}

We have just now seen how strong exponential suppression of the dressed propagator can make the $1$-loop and $2$-loop integrals finite. We believe that most likely this remarkable feature continues to higher loops. The basic reason is - even for the $1$-loop diagrams, the suppression coming from the propagators is stronger than the enhancements coming from the vertices. This ensures two things - first it makes the loops finite, and second the UV growth of the finite diagrams with respect to the external momenta becomes weaker in every subsequent loops. Thus, finiteness of higher loops is  guaranteed recursively. A rigorous proof of the above statement is well beyond the scope of the present paper, but we will now sketch heuristic arguments to demonstrate finiteness of the  particular set of $2$- and $3$-point diagrams that can be constructed out of lower-loop $2$- and $3$-point diagrams, see Fig.~\ref{set}.
%%%%%%%%%%%%%%%%%%%%%%%%%%%%%%%
%%%%%%%%%%%%%%%%%%%%%%%%%%%%%%%%

The basic approach is the following - in order to understand whether any diagram converges in the UV or not, we only need to keep track of the exponential momentum dependences. We already know that the dressed propagators, represented by the shaded blobs, decay in the UV as $e^{-{3\kb^2}/{2}}$. Conservatively, we are therefore going to assume
$\w{\Pi}(k^2)\st{UV}{\longrightarrow} e^{-{3\kb^2}/{2}}$. The $3$-point function (represented by the dark blobs) can, on the other hand, be written as
\be
\label{eq:external}
\Ga_3\st{UV}{\longrightarrow}\sum_{\al,\bt,\ga} e^{\al\pb_{1}^{2}+\bt\pb_{2}^{2}+\ga\pb_{3}^{2}}\ \,,
\ee
with  the convention
\be
\al\geq\bt\geq\ga \,,
\ee
where $p_1,~p_2,~p_3$ are the three external momenta.  This is because once all the (lower-) loop sub diagrams have been integrated out, what remains are expressions in terms of the corresponding external momenta. Some of these external momenta can then become the internal loop momentum in a subsequent higher loop diagram, see Fig.~\ref{set} for a pictorial representation of the recursive construction.

The sum over the exponents $\{\al,~\bt,~\ga\}$ in Eq. (\ref{eq:external}) indicates that there could be many different exponential terms including the permutations needed to symmetrize the vertices over the three internal momenta. We are going to assume that these exponents satisfy certain properties, up to say $(n-1)$-loops. These conditions will allow us to demonstrate that the loops remain finite. Moreover, we will  recursively argue that these properties are also satisfied in
the $n$-th loop.
%%%%%%%%%%%%%%%%
\subsubsection{$2$-point diagram}
\label{sec:2-point}
First, let us look at the zero external momentum limit.
It is easy to see that the most divergent UV part of the $2$-point diagram reads
\be
\label{eq:514}
\Ga_{2,n}{\longrightarrow}\int \frac{d ^ 4 k}{(2 \pi) ^ 4} \, \frac{e^{(\al_1+\al_2+\bt_1+\bt_2)\kb^{2}}}{e^{3\kb^2}}\,,
\ee
where $k$ is the loop momentum variable in Fig.~\ref{set} (left). We've got two propagators $e^{\frac{3 \kb^2}{2}}$   while the (most divergent UV parts of the) vertex factors originating from lower-loop diagrams are $e ^ {\al_{1} \kb^2 + \bt_{1}\kb^2}$ and $e ^ {\al_{2} \kb^2 + \bt_{2}\kb^2}$ (we get no $\ga_{1}$, $\ga_{2}$ terms in the exponents, since the external momenta are set equal to zero). Clearly, the integral is finite as long as
\be
\al_{i}+\bt_{i}<\frac{3}{2}\,,
\label{condition1}
\ee
where $i=1, 2$.  One can check that the same condition ensures finiteness of the diagram even when one includes non-zero external momenta.
%%%%%%%%%%%%%%%%%%%%%
\subsubsection{$3$-point diagram}

First, let us check whether the $3$-point diagram (see Fig.~\ref{set}, right) is finite or not for zero external momenta. Again the most divergent UV contribution comes when the momentum associated with exponents, $\al$'s and $\bt$'s, run in the internal loop giving rise to
\be
\label{eq:516}
\Ga_{3,n}{\longrightarrow}\int \frac{d ^ 4 k}{(2 \pi) ^ 4} \, \frac{e^{(\al_1+\al_2+\al_3+\bt_1+\bt_2+\bt_3)\kb^2} }{e^{\frac{9\kb^2}{2}}}\,,
\ee
where $k$ is the loop momentum variable in Fig.~\ref{set}, right. Similarly to the argument for the $2$-pt function, we've got three propagators $e^{\frac{3 \kb^2}{2}}$,   while the (most divergent UV parts of the) vertex factors originating from lower-loop diagrams are $e ^ {\al_{1} \kb^2 + \bt_{1}\kb^2}$, $e ^ {\al_{2} \kb^2 + \bt_{2}\kb^2}$ and $e ^ {\al_{3} \kb^2 + \bt_{3}\kb^2}$.  Again the integral converges as long as Eq. (\ref{condition1}) is valid.

To prove the validity of Eq. (\ref{condition1}), let us try to find out how one can get the largest exponents for the external momenta. First, let us consider how one can get the largest sum of all the exponents, {\it i.e.}, $\al+\bt+\ga$. Although, all the arguments below can be conducted for three different sets of exponents in the three $3$-point vertices making up the $1$-loop triangle,  Fig.~\ref{set} (right), for simplicity, here we will look at what happens when all the three vertices have the same exponents. Clearly, the best way to obtain the largest exponents for the external momenta is to have the $\al$ exponent correspond to the external momenta. For a symmetric distribution of $(\bt, \ga)$ among the internal loops, we get
\be
\label{eq:crucial}
\Ga_{3,n}{\longrightarrow}\int \frac{d ^ 4 k}{(2 \pi) ^ 4} \frac{e^{\al^{n-1}(\pb_1^2+\pb_2^2+\pb_3^2)}}{e^{[\frac{3}{2}-\bt^{n-1}-\ga^{n-1}][3\kb^2+\frac{1}{3}(\pb_1^2+\pb_2^2+\pb_3^2)]} }\,,
\ee
where $p_{1}$, $p_{2}$, $p_{3}$ are the external momenta for the $1$-loop triangle, and  the superscript in the $\al,\bt,\ga$ indicates that these are coefficients that one obtains from contributions up to $n-1$  loop level.  Before proceeding to obtain the $n$-th loop coefficients, let us briefly explain how we got Eq.~\eqref{eq:crucial}. Assuming symmetrical routing of momenta in the $1$-loop triangle, we get the propagators $e^{-\frac{3}{2} \LF \kb + \frac{\pb_{1}}{3} - \frac{\pb _ {2}}{3} \RF^2}$, $e^{-\frac{3}{2} \LF \kb + \frac{\pb_{2}}{3} - \frac{\pb _ {3}}{3} \RF^2}$ and $e^{-\frac{3}{2} \LF \kb + \frac{\pb_{3}}{3} - \frac{\pb _ {1}}{3} \RF^2}$, and the vertex factors $e ^ {\al^{n-1} \pb _ {1} ^ {2} + \bt^{n-1} \LF \kb + \frac{\pb _ {3}}{3} - \frac{\pb _ {1}}{3} \RF^2 + \ga^{n-1} \LF \kb + \frac{\pb _ {1}}{3} - \frac{\pb _ {2}}{3} \RF ^2}$, $e ^ {\al^{n-1} \pb _ {2} ^ {2} + \bt^{n-1} \LF \kb + \frac{\pb _ {1}}{3} - \frac{\pb _ {2}}{3} \RF^2 + \ga^{n-1} \LF \kb + \frac{\pb _ {2}}{3} - \frac{\pb _ {3}}{3} \RF ^2}$ and $e ^ {\al^{n-1} \pb _ {3} ^ {2} + \bt^{n-1} \LF \kb + \frac{\pb _ {2}}{3} - \frac{\pb _ {3}}{3} \RF^2 + \ga^{n-1} \LF \kb + \frac{\pb _ {3}}{3} - \frac{\pb _ {1}}{3} \RF ^2}$. Conservation of momenta then yields Eq.~\eqref{eq:crucial}.

By integrating Eq. \eqref{eq:crucial}, we have
\be
\al^n=\bt^n=\ga^n=\al^{n-1}+\frac{1}{3}(\bt^{n-1}+\ga^{n-1})-\frac{1}{2} \,.
\label{alpha-n}
\ee
In particular, for the $1$-loop, $3$-point graph, one has to use the $3$-point bare vertices: $\al^0=1$ and $\bt^0=\ga^0=0$. One then obtains
\be
\al^1=\bt^1=\ga^1=\frac{1}{2} \,,
\ee
leading to an overall symmetric vertex: $e^{\frac{1}{2}(\pb_1^2+\pb_2^2+\pb_3^2)}$ and $\al ^ {1}+\bt^{1}+\ga^{1} = \frac{3}{2}$. Since we expect the exponents to decrease as we increase loops, we therefore conjecture that the sum of exponents satisfies the inequality
\be
\al^{n} +\bt^{n} +\ga^{n} \leq \frac{3}{2} \,.
\label{condition2}
\ee
From Eq.~\eqref{alpha-n}, we see that this is satisfied provided a further condition is satisfied by the exponents, \ie:
\be
\al^{n-1}+\frac{1}{3}(\bt^{n-1}+\ga^{n-1})\leq 1 \,.
\label{condition3}
\ee
To summarize, so far we have shown that if, up to $n-1$ loops, inequality Eq.~\eqref{condition3}
is satisfied, then, at the $n$-th loop, Eq.~\eqref{condition2} is also satisfied. To complete the recursive proof, we must argue that Eq.~\eqref{condition3} is also satisfied at $n$-loops. For the loop contribution we are discussing, we have
\be
\al^{n}+\frac{1}{3}(\bt^{n}+\ga^{n})=\frac{5}{3}\LT\al^{n-1}+\frac{1}{3}(\bt^{n-1}+\ga^{n-1})-\frac{1}{2}\RT\leq \frac{5}{6}<1\,,
\ee
and Eq.~\eqref{condition3}  is indeed satisfied.

One may wonder whether there are other ways of distributing the exponents which could violate Eq.~\eqref{condition2}. For instance, one can try to maximize $\al^n$ by distributing
$\al^{n-1}$ in two of the vertices to run along the internal loop. However, one can check that Eq.~\eqref{condition3} still remains valid.

The final point is that the sum of the exponents is maximized by distributing the largest exponents to all the external momentum, thereby ensuring that Eq.~\eqref{condition1} follows from Eq.~\eqref{condition2}. While we do not yet have a rigorous proof of these above arguments, in all the cases we have looked at so far, the inequalities, Eq.~\eqref{condition1}, Eq.~\eqref{condition2} and Eq.~\eqref{condition3} seem to hold up.

%%%%%%%%%%%%%%%%%%%%%%%%%%%%%%%
\section{Summary \& future research directions}\label{sec:conclusions}
\numberwithin{equation}{section}
%%%%%%%%%%%%%%%%%%%%%%%%%%%%%%%

In this paper, we studied the quantum loops for an infinite-derivative scalar field theory action as a toy model to mimic the UV properties of the BGKM
gravity~\cite{Biswas:2011ar}, which is ghost-free at tree-level (see~\cite{Efimov} for a discussion of unitarity in infinite-derivative theories). Expanding the BGKM action around the Minkowski vacuum, one can obtain, for instance,  the ``free'' part that determines the propagator from the $\cO (h^2)$ terms,  while the $\cO (h^3)$ terms determines the cubic interaction vertices. Unfortunately, $\cO (h^3)$  terms  are technically challenging and some of the expression involves double sums. Instead of getting involved with too many technicalities, we therefore chose to work with a simple toy model
action, Eq.~\eqref{eq:action}, that respects a combination of the shift and scaling symmetry at the level of equation of motion that lets us capture some of  the essential features  of BGKM gravity such as the compensating nature of the exponential suppression in the propagator and the exponential enhancement in the  vertex factor.

Of course, in order to formulate a consistent, UV-complete nonlocal theory of quantum gravity, we should discuss the spin-$2$ part of the graviton propagator (the corresponding exponential enhancements from the vertices would also include rank-$2$ tensors). The current paper merely captures some aspects of an infinite-derivative theory of gravity. By no means is our current analysis complete and can only be considered as a step towards formulating a nonlocal infinite-derivative theory of gravity. However, there is a general belief that nonlocality may be able to tame the quantum divergences in gravity that remain one of the outstanding challenges of theoretical physics.

Higher-derivative theories, typically, suffer from the problem of ghosts of the Ostrogradsky instabilities, as discussed quite elaborately by Eliezer \& Woodard in Ref.~\cite{Eliezer:1989cr}. The ghosts typically arise as extra poles in the propagator, while the Ostrogradsky argument relies on having a highest ``momentum'' associated with the highest derivative in the theory in which the energy is seen to be linear, as opposed to quadratic. However, the ``BGKM'' model contains an infinite set of derivatives where no such highest momentum operator can be readily identified, nor are their any extra poles in the propagator which could correspond to new degrees of freedom ghosts or otherwise. This is the reason why we believe that the BGKM theory may be ghost-free. We would like to note that the presence of ghosts typically shows up as classical instabilities. So far, our studies involving certain classical cosmological backgrounds, Ref.~\cite{Biswas:2012bp}, have shown that the perturbations remain under control and do not show any instabilities. Also, we are currently investigating the issue of unitarity more explicitly by looking into scattering processes and revisiting the optical theorem for these theories as there remain certain technical and subtle considerations involving analytic continuation and the use of Cauchy Principal Value theorem.

Even in ordinary field theory, integrals are irregular and one always needs to come up with prescriptions (such as Wick rotation or the $i\en$ prescription) to make sense of the integrals. In other words, quantum field theories have to be supplemented with rules that define what an integral is as the usual rules simply don't work. The same is true for the nonlocal models, in fact, largely our paper can be thought of as an attempt to find prescriptions/defining rules that makes the loop integrals well defined and hopefully (after renormalization of the $1$-loop divergences) finite. All we can say is that, so far, in the loops that we have looked at, our prescription seems to be giving us well defined results which bodes well for future. Whether these prescriptions can be carried forward to higher and higher loops and in the actual theory of gravity obviously remains an open question.

We derived the Feynman rules for our toy model action, {\it i.e.}, the propagator and the vertex factors. Consequently, we computed the $1$-loop, $2$-point diagram, both with zero and arbitrary external momenta, which gives a $\La^4$ divergence, where $\La$ is a momentum cutoff. The $2$-loop diagrams with zero external momenta also give a $\La^4$ divergence, suggesting that we do not get new divergences as we proceed from $1$-loop to $2$-loop. We repeated our
$1$-loop and $2$-loop computations with external momenta, and we paid extra care in understanding the  $1$-loop, $2$-point function which appears as a subdivergence in higher-loop diagrams. Typically, in the $1$-loop, $2$-point function, we obtain a $e^{\frac{3 \pb^2}{2}}$ external momentum dependence in the UV
which indicates that, for $\pb^{2} \ra \infty$, the $1$-loop, $2$-point function tends to infinity. This may appear as an initial setback, but, actually, this external momentum dependence is what, we believe, makes all higher-loop and higher-point diagrams finite once the bare propagators are replaced with the dressed propagators.

This becomes possible because  the exponential suppression in the dressed propagator overcomes the exponential enhancement originating from the vertices. The $1$-loop, $N$-point functions with zero external momenta become UV-finite, as do the $2$-loop integrals for vanishing external momenta. We believe that, even in the case of arbitrary external momenta, our results will not change; the higher-loop diagrams also become UV-finite with the use of the dressed propagators. The basic reason is that, even for the $1$-loop diagrams, the suppression coming from the propagators is stronger than the enhancements coming from the vertices. This ensures two things - first, it makes the loops finite and, second, the UV growth of the finite diagrams with respect to the external momenta becomes weaker in every subsequent loops. Thus, finiteness of higher loops is ensured recursively.

To illustrate this general argument, we considered the finiteness of $n$-loop, $2$-point and $3$-point diagrams that can be constructed out of lower-loop $2$-point and $3$-point diagrams. We found strong arguments, but not rigorous proofs, to support that, once the $1$-loop divergences are tamed, all other higher-loop and higher-point functions may become finite. Whether these results can be transferred to a complete theory of gravity is indeed an open question, but it is a question worth exploring, at the very least. This is already an encouraging sign for an infinite-derivative action of scalar toy model, which can now make higher loops finite, giving us a ray of hope to tackle the problem in full glory for the BGKM gravity. However, as a future computation, it would be interesting to first demonstrate that the finiteness of the diagrams hold to all orders in loops for any $N$-point diagrams. A full proof even for an infinite-derivative  toy model is beyond the scope of this current paper, and we will carry on this computation elsewhere.

\section{Acknowledgments}

Authors would like to thank Damiano Anselmi  for helpful discussions. They would also like to thank the anonymous referee for his/her useful comments. ST is supported by a scholarship from the Onassis Foundation.
AM is supported by the Lancaster-Manchester-Sheffield Consortium for Fundamental Physics under STFC grant ST/J000418/1.
AM is also thankful to the hospitality of Ruth Durrer and the Physics Department of Geneva University where part of the work has been carried out.

\setcounter{equation}{0}
\section{Appendices}
\appendix

%%%%%%%%%%%%%%%%%%%%%%%%%%%%%%%%%%%%%%%%%%%%%%%%%%%%

\section{The BRST-invariant gravitational action and superficial degree of divergence}
\numberwithin{equation}{section}
\label{sec:BRST}

If we consider metric fluctuations around the Minkowski background,
\be
g _ {\mu \nu} = \eta _ {\mu \nu} + h _ {\mu \nu}\,,
\ee
within the harmonic gauge, we can define the quantum theory with:
\be
\p _ {\nu} \bar{h} ^ {\mu \nu} = 0\,,
\ee
where
\be
\bar{h} _ {\mu \nu} = h _ {\mu \nu} - \frac{1}{2} \eta _ {\mu \nu} h \Ra \bar{h} ^ {\mu \nu} = h ^ {\mu \nu}-\frac{1}{2}\eta^{\mu \nu}h \, .
\ee
$\bar{h}_{\mu \nu}$ is called the trace-reverse of $h_{\mu \nu}$ since $\bar{h} = \eta^{\mu \nu} \bar{h}_{\mu \nu} =-h$, where $h = \eta^{\mu \nu} h_{\mu \nu}$. Note that $h ^ {\mu \nu} = \eta ^ {\mu \rho} \eta ^ {\nu \sigma} h _ {\rho \sigma} \Ra \bar{h}^{\mu \nu}=\eta^{\mu \rho} \eta^{\nu \sigma}\bar{h}_{\rho \sigma}$. Furthermore, $h_{\mu \nu}=\bar{h}_{\mu \nu}- \frac{1}{2}\eta_{\mu \nu}\bar{h}$.

The quantized BGKM action must contain:
%{\bf Spyros: Check this, are you really doing for BGKM action or just $R{\cal %F}(\Box) R$?}:
\begin{align}
S _ {\mt{quantized}}& = S + S _ {\mt{GF}} + S _ {\mt{ghost}}\non
& = S_{EH}+S_{Q}+ \frac{1}{2 \xi} \int{d ^ {4} x \, F _ {\tau} \cF (\Box) F ^ {\tau}}+\int{d ^ {4} x \, \bar{C} _ {\tau} \vec{F} _ {\mu \nu} ^ {\tau} D _ {\alpha} ^ {\mu \nu} C ^ {\alpha}}\,.
\end{align}
where $S = S _{EH}+S_Q$ is the gravitational action (see Eq.~\eqref{action}), $S _ {EH}$ is given by Eq.~\eqref{eq:EH} and $S_{Q}$ is given by Eq.~\eqref{nlaction}. $S_{\mt{GF}}$ is the gauge-fixing term and $S_{\mt{ghost}}$ is the ghost-antighost action while $\xi$ is a finite parameter. We have that $F ^ {\tau} = \vec{F} _ {\mu \nu} ^ {\tau} h ^ {\mu \nu}$ and $\vec{F} _ {\mu \nu} ^ {\tau} = \delta _ {\mu} ^ {\tau} \vec{\p}_ {\nu} - \frac{1}{2}\da_{\sigma}^{\tau}\eta^{\sigma \rho}\eta_{\mu \nu}\vec{\p}_{\rho}$ (the arrow indicates the direction in which the derivative acts). $C ^ {\sigma}$ is the ghost field and $\bar{C} _ {\tau}$ is the antighost field; both are anticommuting. $D _ {\al}^{\mu \nu}$ is the operator generating gauge transformations in the graviton field $h^{\mu \nu}$, given an arbitrary infinitesimal vector field $\xi ^ {\alpha} (x)$ (corresponding to $x ^ {' \mu} = x ^ {\mu} - \xi ^ {\mu}$).

That is, $\delta h _ {\mu \nu} = \delta g _ {\mu \nu}= \mathcal{L} _ {\xi} g _ {\mu \nu} = \xi ^ {\rho} \partial _ {\rho} g _ {\mu \nu} + g _ {\mu \rho} \partial _ {\nu} \xi ^ {\rho} + g _ {\rho \nu} \partial _ {\mu} \xi ^ {\rho} = D _ {\mu \nu \alpha} \xi ^ {\alpha}$, where $\mathcal{L}$ is the Lie derivative and
\be
D _ {\mu \nu \alpha} \xi ^ {\alpha} = \partial _ {\mu} \xi _ {\nu} + \partial _ {\nu} \xi _ {\mu} \nonumber +  h _ {\alpha \nu} \partial _ {\mu} \xi ^ {\alpha} + h _ {\mu \alpha} \partial _ {\nu} \xi ^ {\alpha} + \xi ^ {\alpha} \partial _ {\alpha} h _ {\mu \nu}\,.
\ee
Accordingly,
\be
\label{D}
D _{\mu \nu \al} = \eta_{\al \nu} \p_{\mu} + \eta_{\mu \al}\p_{\nu} + h _{\al \nu} \p_{\mu}+ h_{\mu \al}\p_{\nu} + \p_{\al} h _{\mu \nu}\,.
\ee
We can raise indices in~\eqref{D} using the Minkowski tensors.

Moreover, $\cF$ is an analytic function of $\Box$:
\be
\cF (\Box) = \sum _ {n=0}^{\infty} f _ {n} \Box ^n\,,
\ee
where $f_{n}$ are real coefficients.

If we change the gauge-fixing term to read $F^{\tau}=e^{\tau}(x)$~\cite{Stelle:1976gc}, with $e^{\tau}(x)$ an arbitrary $4$-vector function, we can smear out the gauge condition with a weighting functional. Choosing the weighting functional
\be
\omega (e ^ {\tau}) = \exp \LT i\LF\frac{1}{2 \xi} \int{d ^ 4 x \, e _ {\tau} \cF(\Box) e ^ {\tau}}\RF\RT\,,
\ee
where $\xi$ is a finite parameter, we obtain the gauge-fixing term (see~\cite{Tomboulis,Stelle:1976gc} for a derivation of the gauge-fixing term in non-local theories),
\be
S_{\mt{GF}}=  \frac{1}{2 \xi} \int{d ^ {4} x \, F _ {\tau} \cF (\Box) F ^ {\tau}}\,.
\ee
The ghost-antighost action is given by
\begin{comment}
\be
\label{gho}
S_{\mt{ghost}}=\int{d ^ {4} x \, \bar{C} _ {\tau} \, \cF (\Box) \vec{F} _ {\mu \nu} ^ {\tau} D _ {\alpha} ^ {\mu \nu} C ^ {\alpha}}\,.
\ee
If we perform integration by parts $2n$ times for each term in~\eqref{gho} arising from the power series expansion of $\cF(\Box)$,  and redefine appropriately the anti-ghost field, the ghost-antighost action will become:
\end{comment}
\be
S_{\mt{ghost}} = \int{d ^ {4} x \, \bar{C} _ {\tau} \vec{F} _ {\mu \nu} ^ {\tau} D _ {\alpha} ^ {\mu \nu} C ^ {\alpha}}\,,
\ee
following the usual prescription~\cite{Peskin,Weinberg}.

Hereafter, repeated indices denote both summation over the discrete values of the indices and integration over the spacetime arguments of the functions or operators indexed.

The BRST transformations for Yang-Mills theories express a residual symmetry of the effective action which remains after the original gauge invariance has been broken by the addition of the gauge-fixing and ghost action terms~\cite{Stelle:1976gc}. The BRST transformations for BGKM gravity,  appropriate for the gauge-fixing term, $S_{\mt{GF}}$, are given by
\begin{align}
\delta _ {\mt{BRST}} h ^ {\mu \nu} & = D _ {\alpha} ^ {\mu \nu} C ^ {\alpha} \delta \lambda\,, \\
\delta _ {\mt{BRST}} C ^ {\alpha}& = - \partial _ {\beta} C ^ {\alpha} C ^ {\beta} \delta \lambda\,, \\
\delta _ {\mt{BRST}} \bar{C} _ {\tau}& =  \frac{1}{\xi} \, \cF(\Box) F _ {\tau} \delta \lambda\,,
\end{align}
where $\da \la$ is an infinitesimal anticommuting constant parameter. The BRST transformation of the gravitational field is just a gauge transformation of $h ^ {\mu \nu}$ generated by $C ^ {\alpha} \delta \lambda$; thus, gauge-invariant functionals of $h _ {\mu \nu}$, like $S$, are BRST-invariant.
The transformation of $C^{\sigma}$ is nilpotent,
\be
\da_{\mt{BRST}} (\p_{\bt}C^{\sa}C^{\bt})=0\,,
\ee
while the transformation of $h^{\mu \nu}$ is also nilpotent:\be
\da_{\mt{BRST}}(D_{\al}^{\mu \nu}C^{\al}) = 0\,.
\ee
Since the only part of the ghost action which varies under the BRST transformations is the antighost, $\bar{C} _ {\tau}$, we have chosen the BRST transformation of the antighost such that the variation of the ghost action cancels the variation of the gauge-fixing term. Hence, $S_{\mt{quantized}}$ is BRST-invariant.

If we also include BRST-invariant couplings of the ghosts and gravitons to some external fields $K_{\mu \nu}$ (anti-commuting) and $L_{\sigma}$ (commuting),
we obtain the effective action $\widetilde{S}$, as:
\be
\widetilde{S} = S_{\mt{quantized}}+ K _ {\mu \nu} D _ {\al} ^ {\mu \nu} C ^ {\al} + L _ {\sigma} \p _ {\beta} C ^ {\sigma} C ^ {\bt}\,,
\ee
where $\widetilde{S}$ is also BRST-invariant.

Now, we want to compute the superficial degree of divergence for the BRST-invariant BGKM action, where we have made the choices $\cF_3(\Box)=0$ \& $\cF _ {1} (\Box) = \frac{e ^ {- \Box / M^2} - 1}{\Box} = - \frac{\cF _ {2} (\Box)}{2}$ in Eqs.~\eqref{nlaction},~\eqref{quadratic} and $\cF(\Box)= e^{-\Box / M^2}$. To proceed, let us introduce the following notations:
\bi
\item $n_{h}$ is the number of graviton vertices,
\item $n_{G}$ is the number of anti-ghost-graviton-ghost vertices,
\item $n_{K}$ is the number of $K$-graviton-ghost vertices,
\item $n_{L}$ is the number of $L$-ghost-ghost vertices,
\item $I_{h}$ is the number of internal graviton propagators,
\item $I_{G}$ is the number of internal ghost propagators,
\item $E_{C}$ is the number of external ghosts,
\item $E_{\bar{C}}$ is the number of external anti-ghosts.
\ei
By counting the exponential contributions of the propagators and the vertex factors, as discussed in section \ref{intro-1}, we can now obtain the superficial degree of divergence, which is given by
\be
E = n _ {h} - I _ {h}\,,
\ee
where $n _ {h}$ is the number of graviton vertices, and $I _ {h}$ is the number of internal graviton propagators.
By using the following topological relation,
\be
L = 1 + I _ {h} + I _ {G} - n _ {h} - n _ {G} - n _ {K} - n _ {L}\,,
\ee
we get
\be
E = 1 - L + I _ {G} - n _ {G} - n _ {K} - n _ {L}\,.
\ee
Employing the momentum conservation law for ghost and anti-ghost lines,
\be
2 I _ {G} - 2 n _ {G} = 2 n _ {L} + n _ {K} - E _ {C} -E _ {\bar{C}}\,,
\ee
we obtain
\be
E = 1 - L -\frac{1}{2} \left(n _ {K} + E _ {C} + E _ {\bar{C}}\right)\,.
\ee
Note that, as $n_{K}$, $E_C$ and $E_{\bar{C}}$ increase, the degree of divergence decreases. Therefore, the most divergent diagrams are those for which $n_{K}=E_C=E_{\bar{C}}=0$, \ie, the diagrams whose external lines are all gravitons. In this case, the degree of divergence is given by: $E = 1 - L$. For $L \geq 2$, $E<0$ and the corresponding loop amplitudes are superficially convergent.

\section{Types of terms originating from the gravitational action}
\numberwithin{equation}{section}
\label{sec:prototype}

If we compute the $\mathcal{O} (h^3)$ part of the Einstein-Hilbert action, $S_{EH}$ (see Eq.~\eqref{eq:EH}), we obtain the following type of term:
\be
h \p _ {\mu} h \p ^ {\mu} h\,.
\ee
Now, let us compute the $\mathcal{O} (h ^ 3)$ part of the BGKM action, $S _ {Q}$ (see Eq.~\eqref{nlaction}), keeping in mind that $\sqrt{-g} = 1 + \frac{1}{2} h + \frac{1}{8} h ^ {2} - \frac{1}{4} h _ {\nu} ^ {\mu} h _ {\mu} ^ {\nu} +\mathcal{O} (h ^ 3)$, where $h = h _ {\mu} ^ {\mu} = \eta ^ {\mu \nu} h _ {\mu \nu}$.

We shall need the following relation for the double sums appearing in the last two lines of Eq.~\eqref{eq:hello}, see below:
\be
T \delta ( \Box ^ {n})  S = \sum _ {m = 0} ^ {n-1} \Box ^ {m} T \delta(\Box )  \Box ^ {n-m-1} S\,,
\ee
where $n \geq 1$. $\delta \LF \Box \RF$ indicates the variation of the $\Box$ operator and $\delta \LF \Box ^ {n} \RF$ indicates the variation of the $\Box ^ {n}$ operator. $S$ and $T$ are tensors constructed out of the Riemann curvatures and the metric.

Then, by applying integration by parts where appropriate,
\begin{align}
\label{eq:hello}
S _ {Q} ^ {(3)} & = \int \mathrm{d} ^ 4 x \, \frac{1}{2} h \LT R ^ {(1)} \cF _ {1} (\Box) R ^ {(1)} + R _ {\mu \nu} ^ {(1)} \cF _ {2} (\Box) R ^ {(1) \mu \nu} + R _ {\mu \nu \lambda \sigma} ^ {(1)} \cF _ {3} (\Box) R ^ {(1) \mu \nu \lambda \sigma}\RT \non
& + \int \mathrm{d} ^ 4 x \,  \LT \vphantom{ d _ {n} \partial ^ {2m} R _ {\mu \nu \lambda \sigma} ^ {(1)} \Box ^ {(1)} \partial ^ {2(n-m-1)} R ^ {(1) \mu \nu \lambda \sigma}}2 R ^ {(2)} \cF _ {1} (\Box) R ^ {(1)} + R _ {\mu \nu} ^ {(1)} \cF _ {2} (\Box) R ^ {(2) \mu \nu} + R _ {\mu \nu} ^ {(2)} \cF _ {2} (\Box) R ^ {(1) \mu \nu} \Rd \non
& + \Ld R _ {\mu \nu \lambda \sigma} ^ {(1)} \cF _ {3} (\Box) R ^ {(2) \mu \nu \lambda \sigma}+ R _ {\mu \nu \lambda \sigma} ^ {(2)} \cF _ {3} (\Box) R ^ {(1) \mu \nu \lambda \sigma}\RT \non
& + \sum _ {n=1} ^ {\infty} \sum _ {m=0} ^ {n-1} \int \mathrm{d} ^ 4 x \, \LT \vphantom{ d _ {n} \partial ^ {2m} R _ {\mu \nu \lambda \sigma} ^ {(1)} \Box ^ {(1)} \partial ^ {2(n-m-1)} R ^ {(1) \mu \nu \lambda \sigma}} f _ {1_n} \Box ^ {m} R ^ {(1)} \delta(\Box) \Box ^ {n-m-1} R ^ {(1)} + f _ {2_n} \Box ^ {m} R _ {\mu \nu} ^ {(1)} \delta(\Box) \Box ^ {n-m-1} R ^ {(1) \mu \nu} \Rd \non
& + \Ld f _ {3_n} \Box ^ {m} R _ {\mu \nu \lambda \sigma} ^ {(1)} \delta(\Box) \Box ^ {n-m-1} R ^ {(1) \mu \nu \lambda \sigma}\RT\,,
\end{align}
where $\cF _ {i} (\Box) = \sum _ {n=0} ^ {\infty} f _ {i_n} \Box ^n$, $i=1,2,3$, and $\Box = \eta^{\mu \nu}\p_{\mu} \p_{\nu}$. We have that $R ^{\rho}{}_{\sigma \mu \nu}= \p_{\mu}\Ga_{\nu \sigma}^{\rho}-\p_{\nu}\Ga_{\mu \sigma}^{\rho}+\Ga_{\mu \lambda}^{\rho}\Ga_{\nu \sigma}^{\la}-\Ga_{\nu \lambda}^{\rho}\Ga_{\mu \sigma}^{\la}$. Moreover, $R_{\mu \nu}=R^{\sigma} {}_{\mu \sigma \nu}$ and $R= g^{\mu \nu} R _ {\mu \nu}$.

When we lower or raise an index in the Riemann tensors, that is always done with the use of $$g_{\mu \nu} = \eta _ {\mu \nu} + h _ {\mu \nu}$$  and $$g^{\mu \nu} = \eta^{\mu \nu}-h^{\mu \nu}+\dots,$$ respectively. For instance, $R^{(2) \mu \nu} = \eta^{\mu \rho} \eta^ {\nu \sigma} R _ {\rho \sigma}^{(2)} - 2 \eta^{\mu \rho} h^{\nu \sigma} R _{\rho \sigma}^{(1)}$ and $R^{(2)}=\eta^{\mu \nu}R_{\mu \nu}^{(2)}-h^{\mu \nu}R_{\mu \nu}^{(1)}$. Let us mention that
%{\bf Spyros: define $R^{(2)}$ and $R^{(1)}$}.
\begin{align}
R_{\mu \nu \lambda \sigma}^{(1)}&= \frac{1}{2} \LF \p_{\nu}\p_{\lambda}h_{\mu \sigma} + \p_{\mu}\p_{\sigma}h_{\nu \lambda}- \p_{\mu}\p_{\lambda}h_{\nu \sigma}-\p_{\nu}\p_{\sigma}h_{\mu \lambda} \RF \,, \\
R_{\mu \nu} ^{(1)}&= \frac{1}{2}\LF\p_{\sigma}\p_{\mu}h_{\nu}^{\sigma}+\p_{\nu}\p_{\sigma}h_{\mu}^{\sigma}-\p_{\mu}\p_{\nu}h- \Box h_{\mu \nu} \RF \,, \\
R^{(1)} &= \p_{\mu}\p_{\nu}h^{\mu \nu}- \Box h\,.
\end{align}
Following the same method, we can derive the ${\cal O}(h^2)$ expressions for the Riemann tensors.

The terms involving double sums give rise to technical complications when evaluating the Feynman loop integrals. While several $\cF$'s can satisfy Eq.~\eqref{eq:kuku}, following BGKM~\cite{Biswas:2011ar}, we have:
\be
a(\Box) = e ^ {-\Box / M^2} \quad \& \quad \cF _ {3} (\Box) = 0 \Rightarrow  \cF _ {1} (\Box) = \frac{e ^ {- \Box / M^2} - 1}{\Box} = - \frac{\cF _ {2} (\Box)}{2}\,,
\ee
$a(\Box)$ having been defined in Eq.~\eqref{eq:a} and $M$ being a mass scale at which the infinite-derivative modifications become important. Hence, we obtain a ghost-free, infinite-derivative quantum gravitational action, given by Ref.~\cite{Biswas:2011ar,Biswas:2013kla}
\be
\label{eq:QG}
S _ {QG} = \int \mathrm{d} ^ 4 x \, \sqrt{-g} \left \{ \frac{R}{2} + R\LT\frac{e ^ {- \Box / M^2} - 1}{\Box}\RT R - 2 R _ {\mu \nu} \LT\frac{e ^ {- \Box / M^2} - 1}{\Box}\RT R ^ {\mu \nu} \right \} \,.
\ee
Making these assumptions and enforcing a conformal flatness condition, $h_{\mu \nu} =\Omega^2(x) \, \eta_{\mu \nu}$, where $\Omega(x)$ is a smooth, strictly positive function ($\Omega ^ {2} = h/4$ in four-dimensional spacetime), in order to get scalar-type gravitational terms, we obtain the following types of $\mathcal{O}(h^3)$ terms:
\begin{align}
\label{terms}
 &\p _ {\mu} h \p _ {\nu} h \left(\frac{a(\Box) - 1}{\Box}\right) \p ^ {\mu} \p ^ {\nu} h\,, \quad \p _ {\rho} h \p ^ {\rho} h  \left(\frac{a(\Box) - 1}{\Box} \right) \Box h\,, \non
&h \p _ {\mu} \p _ {\nu} h \left(\frac{a(\Box) - 1}{\Box}\right) \p ^ {\mu} \p ^ {\nu} h\,, \quad h \Box h \left(\frac{a(\Box) - 1}{\Box}\right) \Box h\,.
\end{align}
Inspired by the BGKM action~\eqref{eq:QG}, we wish to construct a scalar field theory {\it toy model} that will capture its essential properties and behaviour. After integration by parts, the terms in Eq.~\eqref{terms} which are relevant to the construction of such a scalar field theory {\it toy model}, are
\be
h \p _ {\mu} h \p ^ {\mu} h\,, \quad h \Box h a(\Box) h\,, \quad h \p _ {\mu} h a(\Box) \p ^ {\mu} h\,.
\ee
Therefore, if we choose the free part, $S_{\mt{free}}$, of the action of our scalar field theory {\it toy model} ($\phi$ is the scalar field) of  Eq.~\eqref{free} (so that the propagator shall have an exponential suppression), the interaction part, $S_{\mt{int}}$, will be of the form
\be
S_{\mt{int}} = \frac{1}{M_p} \int \mathrm{d} ^ 4 x \, \left( \al_{1} \phi \partial _ {\mu} \phi \partial ^ {\mu} \phi + \al_{2} \phi \Box \phi a(\Box) \phi +\al_{3} \phi \partial _ {\mu} \phi a(\Box)  \partial ^ {\mu} \phi \right)\,,
\ee
where $\al_1$, $\al_2$ and $\al_3$ are real coefficients.

\section{Arriving at the {\it toy model} action}
\numberwithin{equation}{section}
\label{sec:A}

We start with
\be
S_{\mt{free}} = \frac{1}{2} \int \mathrm{d}^ 4 x \, \LF \phi \Box a(\Box) \phi \RF
\ee
and
\be
S_{\mt{int}} = \frac{1}{M_p} \int \mathrm{d} ^ 4 x \, \LF \al_{1} \phi \partial _ {\mu} \phi \partial ^ {\mu} \phi + \al_{2} \phi \Box \phi a(\Box) \phi +\al_{3} \phi \partial _ {\mu} \phi a(\Box)  \partial ^ {\mu} \phi \RF\,.
\ee
We want the equation of motion of $S_{\mt{scalar}}= S _ {\mt{free}} + S_{\mt{int}}$ to satisfy the following symmetry: $\phi \to (1 + \epsilon) \phi + \epsilon$.
This requirement will fix the values of the coefficients $\al_1$, $\al_2$ and $\al_3$.

After integrating by parts, we can write $S _ {\mt{free}}$ and $S_{\mt{int}}$, as follows:
\be
\label{eq:aside1}
S _ {\mt{free}} =  \int \mathrm{d} ^ 4 x \, \LF \frac{1}{2} \phi \Box \phi + \frac{1}{2} \phi \tilde{a}(\Box) \phi \RF\,,
\ee
\be
\label{eq:aside2}
S_{\mt{int}} = \int \mathrm{d}^4 x \, \LT \LF \al_{1} + \al_{3} -2\al_{2} \RF \phi \partial _ {\mu} \phi \partial ^ {\mu} \phi + \al_{2} \phi \Box \phi \tilde{a}(\Box) \phi +\al_{3} \phi \partial _ {\mu} \phi \tilde{a}(\Box)  \partial ^ {\mu} \phi \RT\,,
\ee
where
\be
\tilde{a} (\Box) = a (\Box) -1\,.
\ee
Hence, $S_{\mt{scalar}}$ can be written as
\begin{align}
\label{lala}
S_{\mt{scalar}} & = \int \mathrm{d}^4 x \, \LF \frac{1}{2}\phi \Box \phi +\frac{1}{M_p}\LF \al_{1} + \al_{3} -2\al_{2} \RF \phi \partial _ {\mu} \phi \partial ^ {\mu} \phi \RF \non
& + \int \mathrm{d}^ 4 x \, \LF \frac{1}{2} \phi \tilde{a}(\Box) \phi +\frac{1}{M_p} \LF \al_{2} \phi \Box \phi \tilde{a}(\Box) \phi +\al_{3} \phi \partial _ {\mu} \phi \tilde{a}(\Box)  \partial ^ {\mu} \phi \RF \RF\, .
\end{align}
Each of the two lines in Eq.~\eqref{lala}, when considered separately, should have invariant equations of motion under the symmetry: $\phi \to (1 + \epsilon) \phi + \epsilon$. Let us write $S_{\mt{scalar}}$ as
\be
S_{\mt{scalar}} = S _ {1} + S _ {2}\,,
\ee
where $S _ {1}$ is the first line in~\eqref{lala}, and $S_{2}$ is the second line in~\eqref{lala}.

If we vary $S _ {1}$, we obtain
\begin{align}
\delta S _ {1} & = \int \mathrm{d} ^ 4 x \, \left(\epsilon \phi \Box \phi + \frac{\epsilon}{2} \Box \phi +\frac{1}{M_p}\LF 3 \epsilon \LF \al_1 + \al_3 - 2 \al_2\RF \phi \partial _ {\mu} \phi \partial ^ {\mu} \phi + \epsilon \LF \al_1 + \al_3  - 2\al_2\RF \partial _ {\mu} \phi \partial ^ {\mu} \phi \right)\RF \non
& = \int \mathrm{d} ^ 4 x \, \left( \epsilon \LF 1 + 2\al_2 -\al_1-\al_3\RF \phi \Box \phi + \frac{3 \epsilon}{M_p} \LF \al_1+\al_3-2\al_2 \RF \phi \partial _ {\mu} \phi \partial ^ {\mu} \phi \right)\,,
\end{align}
up to a total divergence and after integrating by parts, it should be proportional to $S_1$. Therefore, we should have
\be
\label{first}
1 + 2\al_2-\al_1-\al_3 = \frac{3}{2}\,.
\ee
Now, varying $S _ {2}$ yields
\begin{align}
\delta S _ {2} & = \int \mathrm{d} ^ 4 x \, \LF \epsilon \phi \Box \tilde{a} (\Box) \phi + \frac{\epsilon}{2} \Box \tilde{a} (\Box) \phi + \frac{3 \epsilon}{M_p} \al_2 \phi \Box \phi \tilde{a} (\Box) \phi + \epsilon \al_2 \Box \phi \tilde{a} (\Box) \phi \Rd \non
& + \Ld \frac{3 \epsilon}{M_p} \al_3 \phi \partial _ {\mu} \phi \tilde{a} (\Box) \partial ^ {\mu} \phi + \epsilon \al_3 \partial _ {\mu} \phi \tilde{a} (\Box) \partial ^ {\mu} \phi \vphantom{\frac{\epsilon}{2}}\RF \non
&= \int \mathrm{d} ^ 4 x \, \left( \epsilon \LF1 +\al_2 -\al_3 \RF\phi \Box \tilde{a} (\Box) \phi +\frac{1}{M_p}\LF 3 \epsilon \al_2 \phi \Box \phi \tilde{a} (\Box) \phi + 3 \epsilon \al_3 \phi \partial _ {\mu} \phi \tilde{a} (\Box) \partial ^ {\mu} \phi \right)\RF\,,
\end{align}
up to a total divergence and after integration by parts. Again, it should be proportional to the original action, so
\be
\label{second}
1+\al_2-\al_3 = \frac{3}{2}\,.
\ee
From Eqs.~\eqref{first} \& \eqref{second}, we get
\be
\al_1=\al_2=-\al_3=\frac{1}{4}\,.
\ee
As a result,
\be
S_{\mt{int}} = \frac{1}{M_p} \int \mathrm{d} ^ 4 x \, \LF \frac{1}{4} \phi \partial _ {\mu} \phi \partial ^ {\mu} \phi + \frac{1}{4} \phi \Box \phi a(\Box) \phi -\frac{1}{4} \phi \partial _ {\mu} \phi a(\Box)  \partial ^ {\mu} \phi \RF\,.
\ee

\section{Loop integrals}
\numberwithin{equation}{section}

\subsection{$1$-loop integrals with arbitrary external momenta}

\label{sec:D1}

To compute $1$-loop, $2$-point integral with arbitrary external momenta, $p$ and $-p$, we have:
\be
\int \frac{\mathrm{d}^4 k}{(2 \pi)^4} \, f(p,k)\,,
\ee
where $f$ is a function of the external momentum $p$, and the loop momentum $k$. We analytically continue the integrand, so that we can work in the Euclidean space,
\be
\mathrm{d}^4 k\to i \, \mathrm{d}^4 k\,,~~~~k ^ 2 \to k _ {E}^2\,,~~~~p \cdot k \to (p \cdot k)_{E}\,,~~~~p ^ 2 \to p _ {E} ^ 2\,,
\ee
since $k _ {0} \to i k _ {0}$, and  $p _ {0} \to i p _ {0}$~\footnote{In Minkowski space (mostly plus metric signature), $k^2 = - k_{0}^2 + \vec{k}^2$, where $\vec{k}^2 = k_{1}^{2}+k_{2}^{2}+k_{3}^{2}$. After analytic continuation, $k_{E}^2 = k_{4}^2 + \vec{k}^2$, where $k_{4} = -i k_{0}$.}.
Then, by spherical symmetry, we express $\mathrm{d}^ 4 k_{E}$, as
\be
4 \pi k _{E} ^ {3} \sqrt{1-x^2}\, \mathrm{d}x \, \mathrm{d k}_{E}\,,
\ee
where $x$ is the cosine of the angle between $p_{E}$ and $k _ {E}$. We assume $p _ {E}$ to be the $z$-axis, so $p _ {E} \cdot k _ {E} =  \mathrm{p}_{E} \, \mathrm{k}_{E} \, x$ and $\mathrm{p}_ {E}$ \& $\mathrm{k}_ {E}$ are
the norms of $p_{E}$ and $k _ {E}$.

We then integrate with respect to $x$ from $-1$ to $1$. If the integral converges, we subsequently integrate with respect to $\mathrm{k}_{E}$ from $0$ to $\infty$. If the integral diverges, we integrate with respect to $\mathrm{k}_{E}$ from $0$ to $\Lambda$, where $\Lambda$ is the momentum cutoff.

\subsection{$2$-loop integrals with zero external momenta}

\label{sec:D2}

%See the Mathematica notebook file 21-May-2014.nb.

Let us compute the integral resulting in Eq.~\eqref{eq:imaginary}. The integral is given by:
\be
\label{eq:D4}
\int \frac{\mathrm{d} ^ 4 k _ {1}}{(2 \pi) ^ 4} \frac{\mathrm{d} ^ 4 k _ {2}}{(2 \pi) ^ 4} \, \frac{(k _ {1} ^ {2} + k _ {2} ^ {2} + k _ {3} ^ {2}) ^ {2}}{16 M _ {p} ^ {4} k _ {1} ^ {2} k _ {2} ^ {2} k _ {3} ^ {2} } e ^ {k _ {1} ^ {2} / 2 M ^ 2} e ^ {- (k _ {2} - k _ {3}) ^ {2} / 2 M ^ {2}}\,.
\ee
This can be written as
\be
\label{eq:23}
\frac{1}{16 M _ {p} ^ {4}} \int _ {0} ^ {\La} \mathrm{d} a \int _ {0} ^ {\infty} \mathrm{d} b \int _ {-1} ^ {1} \mathrm{d} x \, \frac{\sqrt{1-x^2} \left(4 \pi  a ^ 3 2 \pi ^2 b^3\right) \left(\frac{3 a^2}{2} + \frac{b^2}{2}\right)^2 \exp \left(\frac{a^2}{2 M^2}\right) \exp \left(-\frac{b^2}{2 M^2}\right)}{16 (2 \pi) ^ 8 a ^ {2} \frac{1}{4} \left(a ^ 2 + b ^ 2 + 2 a b x\right) \frac{1}{4} \left(a ^ 2 + b ^ 2 - 2 a b x\right)}\,,
\ee
where $x$ is the cosine of the angle between $k _ {1}$ and $k _ {2} - k _ {3}$, $a$ is the norm of $k _ {1}$ in the Euclidean space, and $b$ is the norm of $k _ {2}  - k _ {3}$ in the Euclidean space. The factor ${1}/{16}$ in front of the integral is the Jacobian $\left(\frac{1}{2} \right) ^ 4$.

Integrating with respect to $x$ from $-1$ to $1$, we get
\be
\frac{1}{2048 \pi^{5} M _ {p} ^ {4}} \int _ {0} ^ {\La} \mathrm{d} a \int _ {0} ^ {\infty} \mathrm{d} b \, \frac{\pi  \left(a^2+b^2-(a+b) \left| a-b\right| \right)}{4 a^2 b^2 \left(a^2+b^2\right)} a b^{3} \left(3 a^{2}+b^{2} \right)^{2} \exp \left(\frac{a^2}{2 M^2}\right) \exp \left(-\frac{b^2}{2 M^2}\right).
\ee
The integration with respect to $b$ is split into two parts: i) from $0$ to $a$  and ii) from $a$ to $\infty$. The first part becomes ($a^2+b^2-(a+b)(a-b)=2b^2$)
\be
\frac{1}{4096 \pi^{4} M _ {p} ^ {4}} \int _ {0} ^ {\La} \mathrm{d} a \int _ {0} ^ {a} \mathrm{d} b \, \frac{b^3}{a(a^2+b^2)} \left(3 a^{2}+b^{2} \right)^{2} \exp \left(\frac{a^2}{2 M^2}\right) \exp \left(-\frac{b^2}{2 M^2}\right)
\ee
and gives
\begin{align}
\label{eq:first}
\frac{1}{4096 \pi^4 M _ {p} ^ {4}} & \int _ {0} ^ {\La} \mathrm{d} a \, \frac{-2}{a} \left(M^2 \left(a^2 M^2 \left(7-5 e^{\frac{a^2}{2M^2}}\right)-4 M^4 \left(e^{\frac{a^2}{2M^2}}-1\right)+a^4 \left(5-2 e^{\frac{a^2}{2M^2}}\right)\right) \Rd \non
& \Ld + a^6 e^{\frac{a^2}{M^2}} Ei\left(-\frac{a^2}{M^2}\right)- a^6 e^{\frac{a^2}{M^2}} Ei\left(-\frac{a^2}{2M^2}\right) \right).
\end{align}
The second part becomes ($a^2+b^2-(a+b)(b-a)=2a^2$)
\be
\frac{1}{4096 \pi^{4} M _ {p} ^ {4}} \int _ {0} ^ {\La} \mathrm{d} a \int _ {a} ^ {\infty} \mathrm{d} b \, \frac{a b}{a^2+b^2} \left(3 a^{2}+b^{2} \right)^{2} \exp \left(\frac{a^2}{2 M^2}\right) \exp \left(-\frac{b^2}{2 M^2}\right)
\ee
and gives
\be
\label{eq:second}
\frac{1}{4096 \pi^4 M _ {p} ^ {4}} \int _ {0} ^ {\La} \mathrm{d} a \, \frac{1}{4} a \left(24 a^2 M^2+ 8 M^{4}-8 a^4 e^{\frac{ a^2}{M^2}} Ei\left(-\frac{a^2}{M^2}\right)\right).
\ee

In~\eqref{eq:first}, there are (four) terms which diverge exponentially; those terms are the terms in the integrand involving $-5e^{a^2/2M^2}$, $e^{a^2/2M^2}-1$, $-2 e^{a^2/2M^2}$ and $-a^6 e^{a^2/M^2}Ei(-a^2/2M^2)$. For those terms, we first write $M^2=-\tilde{M}^{2}$ and then analytically continue back the integrals to obtain them as a function of  $M^2$~\footnote{We note that the loop integrals in the type of nonlocal theories we are considering are always ill-defined in Minkowski space-time; this is because a term such as $e^{-p^2/M^2}=e^{p_{0}^{2}/M^2}e^{-p_{i}^{2}/M^2}$ is always divergent either in the space or the time direction depending upon the sign of $M^2$. Thus, these integrals only make sense  once appropriately ``Euclideanized" and then analytically continued back to Minkowski space-time. While going to Euclidean space, we always have a choice, either $t\ra it$, or $x\ra ix$, and that depends on the overall sign of the exponents. When we write $M^2=-\tilde{M}^2$ in appendix~\ref{sec:D2}, we simply mean that rather than $t\ra it$, we choose $x\ra ix$. And, as we have now checked, once the integrals are evaluated correctly, we do not find any inconsistencies, all the loop amplitudes are purely imaginary leading to terms that are real in the effective action. Perhaps it is also worth pointing out that the analytic continuations followed here is not new, and has been used in previous nonlocal quantum field theory literature with consistent results. In particular, in both~\cite{Biswas:2014yia,Reddy}, it was also shown that, in the $M^2\ra\infty$ limit, one recovers the local field theory results, as it should.}. We should mention that we get $\Lambda^4$, $\Lambda^2$ and $\log \left( \frac{\Lambda}{M} \right)$ divergences after we apply that prescription in~\eqref{eq:first}. The full result is
\begin{align}
\label{dd9}
\frac{M^2}{8192 \pi^4 M_{p}^{4}} & \left(-4 \Lambda ^4-4 (5+\gamma ) M^4-18 \Lambda ^2 M^2 -2 e^{\frac{\Lambda ^2}{M^2}} \left(\Lambda ^4-2 \Lambda ^2 M^2+2M^4\right) Ei\left(-\frac{\Lambda ^2}{M^2}\right)\Rd \non
&  +8 M^4 \left(\frac{1}{2}Ei\left(-\frac{\Lambda ^2}{2 M^2}\right)+\log \left(\frac{2 M}{\Lambda }\right)\right)+20 M^4 e^{-\frac{\Lambda ^2}{2 M^2}}-4 \Lambda ^2 M^2 e^{-\frac{\Lambda ^2}{2 M^2}}\non
& \Ld +4 \Lambda ^2 M^2 e^{-\frac{\Lambda ^2}{M^2}} Ei \left(\frac{\Lambda ^2}{2 M^2}\right)+2 \Lambda ^4 e^{-\frac{\Lambda ^2}{M^2}} Ei \left(\frac{\Lambda ^2}{2 M^2}\right)+4 M^4 e^{-\frac{\Lambda ^2}{M^2}} Ei \left(\frac{\Lambda ^2}{2 M^2}\right) \right).
\end{align}

Integrating \eqref{eq:second} with respect to $a$ from $0$ to $\Lambda$ yields
\begin{align}
\label{dd10}
\frac{M^2}{8192 \pi^{4} M_{p}^{4}}&\left(-2 e^{\frac{\Lambda ^2}{M^2}} \left(\Lambda ^4-2\La^2 M^2+2 M^4\right) Ei\left(-\frac{\Lambda ^2}{M^2}\right)+4 \Lambda ^4+8 M^4 \log \left(\frac{\Lambda}{M} \right)\Rd \non
 & \Ld + 4 \gamma M^{4}-2\La^2 M^2\vphantom{\mathrm{Ei}\left(-\frac{\Lambda ^2}{M^2}\right)}\right).
\end{align}
%It diverges as $\Lambda^4$, using the relation
%\begin{equation}
%\lim _ {x \to + \infty} x ^ {2} e ^ {\alpha x ^ {2}} Ei \left(- \alpha x %^ {2} \right) = - \frac{1}{\alpha}.
%\end{equation}

Summing \eqref{dd9} and \eqref{dd10}, we obtain
\begin{align}
\label{d13}
\frac{M^2}{2048 \pi^{4} M_{p}^{4}} & \left(-e^{\frac{\Lambda ^2}{M^2}} \left(\Lambda ^4-2\La^2 M^2+2 M^4\right) Ei\left(-\frac{\Lambda ^2}{M^2}\right)-5 M^2 \left(\Lambda ^2+M^2\right) \Rd \non
& +2 M^4 \left(\frac{1}{2}Ei\left(-\frac{\Lambda ^2}{2M^{2}}\right)+\log (2)\right)+5 M^4 e^{-\frac{\Lambda ^2}{2 M^2}}- \Lambda ^2 M^2 e^{-\frac{\Lambda ^2}{2 M^2}}\non
& \Ld + \Lambda ^2 M^2 e^{-\frac{\Lambda ^2}{M^2}} Ei \left(\frac{\Lambda ^2}{2 M^2}\right)+\frac{1}{2} \Lambda ^4 e^{-\frac{\Lambda ^2}{M^2}} Ei \left(\frac{\Lambda ^2}{2 M^2}\right)+ M^4 e^{-\frac{\Lambda ^2}{M^2}} Ei \left(\frac{\Lambda ^2}{2 M^2}\right) \right).
\end{align}

If we expand~\eqref{d13} for large $\La$, 
%(use the Mathematica Series command in $\La$ about $\infty$ for the first %term in~\eqref{d13}), where $x=\La/M$, 
%\be
%Series\left[-\left(x^4-2 x^2+2\right) \exp \left(x^2\right) Ei\left(-x^2\right),\{x,\infty %,10\}\right]= x^2-3+\frac{6}{x^2}+\dots
%\ee
%and take the $\Lambda \ra \infty$ limit, 
we get a quadratic divergence as expected: 
\be
\frac{M^4}{2048 \pi^{4} M_{p}^{4}} \left(M^2 \left(\log (4)-8\right)-4 \Lambda ^2 \right).
\ee

\section{Dimensional regularization}
\numberwithin{equation}{section}

\subsection{Overview of \textbf{DR}}

\label{sec:DR}

If we want to regulate dimensionally an integral, we should follow a certain procedure, see Refs.~\cite{DR,Schwartz,Osborn}. First, we make the following replacement:
\be
\int \frac{d^4 k}{(2 \pi) ^ 4} \,  \frac{g(p,k)}{h(p,k)} \to \int \frac{d^{d} k}{(2 \pi) ^ d} \frac{g(p,k)}{h(p,k)}\,,
\ee
as we now wish to perform the integral in $d$ dimensions, where $d$ is an arbitrary complex number. Then we express the terms appearing in the denominator $h(p,k)$ as integrals, using the following relation for positive $x$:
\be
\frac{1}{x} = \int_{0}^{\infty} d \alpha \, e ^ {- \alpha x}\,;
\ee
$\al$ is called a Schwinger parameter. We complete the square in the exponent of the integrand and then shift the loop momentum variable, so we just have to perform a Gaussian integral. Regarding the numerator $g(p,k)$, we accordingly shift the loop momentum variable (the integration measure is invariant) in order to be consistent, and drop terms linear in the (shifted) $k$ as they integrate symmetrically to $0$. In the context of dimensional regularization, we can also make the replacement $k ^ {\mu} k ^ {\nu} \ra \frac{\delta ^ {\mu \nu} k ^ 2}{d}$, when evaluating the loop integrals (in Euclidean space, after analytic continuation).

For the sake of convenience, let us list the following generalized Gaussian integrals ($a > 0$) in $d$-dimensional Euclidean space, where $d$ is a complex parameter~\cite{DR}:
\be
\label{eq:e1}
\int \frac{d ^ {d} k}{(2 \pi) ^ {d}} \, \exp \left[- a k ^ {2}\right] = \frac{1}{(4 \pi a) ^ {d/2}}\,,
\ee
\be
\int \frac{d ^ {d} k}{(2 \pi) ^ {d}} \, k _ {\mu} \exp \left[- a k ^ {2}\right] = 0\,,
\ee
\be
\int \frac{d ^ {d} k}{(2 \pi) ^ {d}} \, k _ {\mu} k _ {\nu} \exp \left[- a k ^ {2}\right] = \frac{\delta _ {\mu \nu}}{2 a (4 \pi a) ^ {d/2}}\,,
\ee
\be
\int \frac{d ^ {d} k}{(2 \pi) ^ {d}} \, k ^ {2} \exp \left[- a k ^ {2}\right] = \frac{d}{2 a (4 \pi a) ^ {d/2}}\,,
\ee
\be
\int \frac{d ^ {d} k}{(2 \pi) ^ {d}} \, k _ {\mu} k _ {\nu} k _ {\rho} \exp \left[- a k ^ {2}\right] = 0\,,
\ee
\be
\int \frac{d ^ {d} k}{(2 \pi) ^ {d}} \, k _ {\mu} k _ {\nu} k _ {\rho} k _ {\sigma} \exp \left[- a k ^ {2}\right] = \frac{\delta _ {\mu \nu} \delta _ {\rho \sigma} + \delta _ {\mu \rho} \delta _ {\nu \sigma} + \delta _ {\mu \sigma} \delta _ {\nu \rho}}{4 a ^ {2} (4 \pi a) ^ {d/2}}\,,
\ee
\be
\int \frac{d ^ {d} k}{(2 \pi) ^ {d}} \, k ^ {4} \exp \left[- a k ^ {2}\right] = \frac{d (d+2)}{4 a ^ {2} (4 \pi a) ^ {d/2}}\,,
\ee
where $k _ {\mu}$ is a vector in $d$-dimensional Euclidean space. Partial differentiation of \eqref{eq:e1} with respect to $a$ yields the other formulae. All integrals involving odd powers of $k$ vanish.

After the Gaussian integration, there are only some parameter integrals remaining. For instance, if there are two parameter integrals remaining $\int _ {0} ^ {\infty} \mathrm{d} \alpha _ {1} \int _ {0} ^ {\infty} \mathrm{d} \alpha _ {2}$, we can make the following substitutions:
\be
\alpha _ {1} + \alpha _ {2} = s\,, \quad \alpha _ {1} = s \alpha\,, \quad \alpha _ {2} = s (1 - \alpha)\,,
\ee
where $0<s<\infty$ and $0<\alpha<1$ while $d\alpha _{1}d\alpha_{2}=s \, ds \, d\alpha$.

If the integral we are trying to regulate dimensionally is very complicated, we may follow an alternative procedure. We can employ the following relation for the volume element of a $d-1$-dimensional surface:
\be
\label{eq:omega}
\mathrm{d} \Omega _ {d} = \mathrm{d} \Omega _ {d-1} \LF 1 - z^2 \RF ^ {\frac{d - 3}{2}} \mathrm{d} z\,,
\ee
where $z \equiv \cos(\theta)$ ($\theta$ may be defined to be the angle between $p$ and $k$ in $1$-loop Feynman integrals with arbitrary external momenta or the angle between $k _ {1}$ and $k _ {2}$ in $2$-loop Feynman integrals with the external momenta set equal to zero) and $\mathrm{d} \Omega _ {d}$ denotes the differential solid angle of the $d$-dimensional unit sphere:
\be
\mathrm{d} \Omega _ {d} = \sin^{d-2}(\phi _ {d-1})\sin ^{d-3}(\phi _ {d-2})\cdot \cdot \cdot\sin(\phi_2) \mathrm{d} \phi_1 \cdot \cdot \cdot \mathrm{d} \phi _ {d-1}\,,
\ee
where $\phi _ {i}$ is the angle to the $i$-th axis, with $0 \leq \phi _{1}<2\pi$ and $0\leq\phi_{i}<\pi$ for $i > 1$~\cite{Schwartz}.

Then we can use
\be
\int \mathrm{d} ^ d k = \int \mathrm{d} \Omega _ {d} \int k ^ {d-1} \mathrm{d}  k\,.
\ee
and, subsequently, insert Eq.~\eqref{eq:omega}, for which
\be
\Omega _ {d-1} = \int \mathrm{d} \Omega _ {d-1} = \frac{2 \pi ^ {\frac{(d-1)}{2}}}{\Gamma\LF \frac{d-1}{2} \RF}\,;
\ee
$\Gamma(x)$ is the Gamma function.

After all the Gaussian integrals and as many as possible of the parameter integrals have been carried out, we write $d$ as $d = 4 - \epsilon$ and, then, perform a series expansion in $\epsilon$ about $0$. To make the result dimensionally correct, we may have to multiply it by factors of $M^{\epsilon}$, where $M$ is a mass scale at which the non-local modifications become important. If we get a pole in $\epsilon$, say $\frac{1}{\epsilon}$, this means we have a divergence. Otherwise, the integral is finite within the framework of dimensional regularization; this does not necessarily mean that the integral is convergent in the conventional sense.

\subsection{$e^{2 \pb \cdot \kb}$ integrals}

\label{sec:DR2}

The eighth and the ninth terms in Eq.~\eqref{Vextsq}, \textit{i.e.},
\be
\LF e ^ {2\LF\frac{\pb}{2} + \kb\RF ^ {2}}-e ^ {\LF\frac{\pb}{2} + \kb\RF ^ {2}} e ^ {\LF\frac{\pb}{2} - \kb\RF ^ {2}}\RF
\ee
and
\be
\LF 
e ^ {2\LF\frac{\pb}{2} - \kb\RF ^ {2}}-e ^ {\LF\frac{\pb}{2} + \kb\RF ^ {2}} e ^ {\LF\frac{\pb}{2} - \kb\RF ^ {2}}\RF \,,
\ee
give rise to integrals containing the terms $e^{2 \pb \cdot \kb}-1$ and $e^{-2 \pb \cdot \kb}-1$, respectively. Dimensionally regularizing, the sum of the integrals associated with the terms $e^{2 \pb \cdot \kb}-1$ and $e^{-2 \pb \cdot \kb}-1$ is equal to $0$. That is, we refer to the integrals
\be
Q_{1}=\frac{i}{2 M _ {p} ^ {2}} \int \frac{d ^ {4} k}{(2 \pi) ^ {4}} \, \frac{C^{2}( e^{2 \pb \cdot \kb}-1 )}{(\frac{p}{2} + k) ^ {2} (\frac{p}{2} - k) ^ {2}} \ee
and
\be
Q_{2}=\frac{i}{2 M _ {p} ^ {2}} \int \frac{d ^ {4} k}{(2 \pi) ^ {4}} \, \frac{C^{2}( e^{-2 \pb \cdot \kb}-1 )}{(\frac{p}{2} + k) ^ {2} (\frac{p}{2} - k) ^ {2}} \,,
\ee
where
\be
C = \frac{1}{4} \LT p ^ {2} + \LF \frac{p}{2} + k\RF ^ {2} + \LF \frac{p}{2} - k\RF ^ {2}\RT \,.
\ee
Using a Taylor series expansion, one obtains
\be
e^{2 \pb \cdot \kb}-1= \sum_{m=1}^{\infty} \frac{(2\pb \cdot \kb)^{m}}{m!}
\ee
and
\be
e^{-2 \pb \cdot \kb}-1= \sum_{m=1}^{\infty} \frac{(-2\pb \cdot \kb)^{m}}{m!} \,.
\ee 
Therefore, we have that
\be 
\label{eq:lulo}
(e^{2 \pb \cdot \kb}-1)+(e^{-2 \pb \cdot \kb}-1)=2 \sum_{m=1}^{\infty} \frac{(2\pb \cdot \kb)^{2m}}{(2m)!} \,.
\ee
Each of the terms in the following sum,
\be
Q_{1}+Q_{2}=\sum_{m=1}^{\infty} \frac{i}{M _ {p} ^ {2}} \int \frac{d ^ {4} k}{(2 \pi) ^ {4}} \, \frac{C^{2}(2\pb \cdot \kb)^{2m}}{(\frac{p}{2} + k) ^ {2} (\frac{p}{2} - k) ^ {2}(2m)!} \,,
\ee
is equal to zero within the framework of dimensional regularization. To elaborate, we want to compute the integrals of the form
\be
\frac{i}{M _ {p} ^ {2}} \int \frac{d ^ {d} k}{(2 \pi) ^ {d}} \, \frac{C^{2}(2\pb \cdot \kb)^{2m}}{(\frac{p}{2} + k) ^ {2} (\frac{p}{2} - k) ^ {2}(2m)!} \,,
\ee
where $m=1,\dots,\infty$ and $d$ is an arbitrary complex number. First, we Schwinger-parameterize the terms in the denominator
\be
\frac{1}{(\frac{\pb}{2} + \kb) ^ {2}} = \int_{0}^{\infty} d \alpha_{1} \, e ^ {- \alpha_{1} (\frac{\pb}{2} + \kb) ^ {2}} \,,
\ee
\be
\frac{1}{(\frac{\pb}{2} - \kb) ^ {2}} = \int_{0}^{\infty} d \alpha_{2} \, e ^ {- \alpha_{2} (\frac{\pb}{2} - \kb) ^ {2}}\,.
\ee
Then we complete the square in the exponent of the integrand and shift the loop momentum variable suitably. Consequently, we evaluate the Gaussian integrals in Euclidean space and make the following change of variables for the parameter integrals with respect to $a_1$ and $a_2$:
\be
\alpha _ {1} + \alpha _ {2} = s\,, \quad \alpha _ {1} = s \alpha\,, \quad \alpha _ {2} = s (1 - \alpha)\,,
\ee
where $0<s<\infty$, $0<\alpha<1$ and $d\alpha _{1}d\alpha_{2}=s \, ds \, d\alpha$. Finally, we write $d$ as $d = 4 - \epsilon$ and perform a series expansion in $\epsilon$ about $0$. We get no pole in $\en$ and the finite part of the integral is equal to zero. That is, dimensionally regularizing, we obtain
\be
\frac{i}{M _ {p} ^ {2}} \int \frac{d ^ {d} k}{(2 \pi) ^ {d}} \, \frac{C^{2}(2\pb \cdot \kb)^{2m}}{(\frac{p}{2} + k) ^ {2} (\frac{p}{2} - k) ^ {2}(2m)!} =0 \,.
\ee
Therefore,
\be
\Ga_{2,1,iii}(p^2)=Q_{1}+Q_{2}=0 \,.
\ee

%%%%%%%%%%%%%%%%%%%%%%%%%%%%

\end{document}